\documentclass[a4paper,11pt]{article}
\pdfoutput=1 

\usepackage{jcappub} 

\usepackage[T1]{fontenc} 

\usepackage[margin=0.9in]{geometry}
\usepackage{cleveref}
\usepackage{natbib}
\usepackage{amssymb}
\usepackage{amsmath}
\usepackage{siunitx}
\usepackage{physics}
\usepackage{bm}
\usepackage[normalem]{ulem}
\usepackage{graphicx,xcolor}
\usepackage{booktabs}
\usepackage{comment}
\usepackage{stackengine}
\newcommand\xrowht[2][0]{\addstackgap[.5\dimexpr#2\relax]{\vphantom{#1}}}
\usepackage{accents}
\usepackage{trimclip}
\DeclareRobustCommand*{\phat}[1]{{\accentset{(\!\trimbox{0pt 1.1ex}{\ensuremath{\string^}}\!)}{#1}}}
\usepackage{adjustbox}
\usepackage{multirow}

\newcommand{\AI}[1]{{\color{blue} #1}}
\newcommand{\SP}[1]{{\color{orange} #1}}
\newcommand{\JS}[1]{{\color{red} #1}}

\title{\Huge Gravitational Waves from Primordial Black Hole Evaporation with Large Extra Dimensions}

\author[a]{Aurora Ireland,}
\author[b,c]{Stefano Profumo,}
\author[b,c]{and Jordan Scharnhorst}
\affiliation[a]{
Department of Physics, University of Chicago, 5720 South Ellis Avenue, Chicago, IL~60637, USA
}
\affiliation[b]{
	Department of Physics, University of California, Santa Cruz,
	1156 High Street, Santa Cruz, CA 95064, USA
}
\affiliation[c]{
	Santa Cruz Institute for Particle Physics,
	1156 High Street, Santa Cruz, CA 95064, USA
}

\emailAdd{anireland@uchicago.edu}
\emailAdd{profumo@ucsc.edu}
\emailAdd{jscharnh@ucsc.edu}

\abstract{

The spectra of gravitational waves from black hole evaporation generically peak at frequencies of order the Hawking temperature, making this signal ultra-high frequency for primordial black holes evaporating in the early universe. This motivates us to consider small black holes in theories with large extra dimensions, for which the peak frequency can be lowered substantially, since the true bulk Planck scale $M_*$ can be much smaller than the effective $M_{\rm Pl}$. We study the emission of brane-localized gravitons during the Hawking evaporation of ultra-light primordial black holes in the context of theories with large extra dimensions, with the ultimate goal of computing the contribution to the stochastic gravitational wave background. To accurately model black hole evolution, we compute greybody factors for particles of spin-0, 1/2, 1, and 2 emitted on the brane and in the bulk, presuming the majority of emission proceeds during the Schwarzschild phase. We then compute the power spectrum and present day spectral density parameter for brane-localized gravitons contributing to a gravitational wave signal. We find that for an optimal choice of parameters, the peak frequency plateaus in the sub-MHz regime, within range of planned high-frequency gravitational wave detectors, making this scenario a target for detection once their sensitivity exceeds $\Delta N_{\rm eff}$ bounds.
}

\begin{document}
\maketitle
\flushbottom

\section{Introduction}
\label{sec:introduction}

Theories with large extra dimensions (LED) have long been appreciated as a solution to the hierarchy problem \cite{Argyres:1998qn,Arkani-Hamed:1998cxo,Arkani-Hamed:1998jmv}. In such models, the conventional 4-dimensional Planck scale $M_{\rm Pl}$ is merely an effective energy scale related to the true fundamental quantum gravity scale $M_*$ via the relation
\begin{equation}
    M_{\rm Pl}^2 \sim  R^n M_*^{n+2} \,,
\end{equation}
where $R$ is the characteristic size of one of the $n$ spacelike, compact extra dimensions. Clearly for large $R \gg \ell_{\rm Pl}$, $M_*$ can be lowered significantly, such that gravity's apparent weakness can be understood as simply an artifact of the presence of these $n$ large extra dimensions. In order that the other fundamental forces are not affected beyond phenomenological limits, the particle content of the Standard Model must be confined to a 4-dimensional ``brane'' embedded in this $(n+4)$-dimensional ``bulk'', in which only gravitons and possibly heavy Kaluza-Klein scalar modes can propagate. See e.g. \cite{Rattazzi:2003ea} for a pedagogical review. Presuming a fundamental Planck scale $M_* \simeq 10\ \text{TeV}$, the extra dimensions must be excessively large for $n=1$ , with $R \sim 10^{12} \text{cm}$ --- in conflict with precision tests of Newton's law. Larger values of $n$ are subject to constraints from various cosmological and astrophysical sources as well as collider searches, as we will review, but remain phenomenologically viable and interesting \cite{Workman:2022ynf}. 

The dimensionality of spacetime generically affects solutions in general relativity, including black hole solutions. Black holes in LED scenarios are centered on the brane but extend along the extra dimensions. Because the fundamental Planck scale is lowered, black holes can also be produced in energetic particle collisions in the hot thermal plasma of the early universe, in addition to the usual formation mechanisms. Small black holes with horizon radius less than the characteristic scale of the extra dimensions $r_h \ll R$ are ($n+4$)-dimensional objects, while larger black holes with $r_h \gg R$ behave as effectively 4-dimensional. In either case, black hole properties --- notably temperature and the spectrum of Hawking radiation --- are affected by the presence of these extra dimensions. 

In this paper, we investigate the gravitational wave signal from the evaporation of ultra-light primordial black holes in the context of theories with LED. The spectrum of gravitational waves from black hole evaporation peaks at a frequency which is an order one factor times the Hawking temperature, $\omega_{\rm peak} \sim T_H$. In the 4-dimensional case then, $\omega_{\rm peak} \sim M_{\rm Pl}^2/M$, which is generically quite high for primordial black holes with $M \lesssim 5 \times 10^8\, \text{g}$, as required for evaporation before big bang nucleosynthesis (BBN). It was shown in Ref.~\cite{Ireland:2023, Dong_2016} that even after redshifting these signals to the present day, the peak frequencies tend to be outside of the range of proposed ultra-high frequency gravitational wave detection technologies. This motivates us to consider black holes in extra dimensional scenarios, which are often colder than their 4-dimensional counterparts of the same mass. As we will see, the peak frequency in this case scales with black hole mass $M$ as 
\begin{equation}
    \omega_{\rm peak} \sim \left( \frac{M_*}{M} \right)^{\frac{1}{n+1}} M_* \,.
\end{equation}
Since the true quantum gravity scale $M_*$ can be much lower than the observed $M_{\rm Pl}$, we expect $\omega_{\rm peak}$ to be much lower --- perhaps even within reach of future gravitational wave detectors. 

This study is structured as follows: the next section briefly reviews black holes in the LED setup we consider, as well as general constraints on LED scenarios; Sec.~\ref{sec:edpbh} reviews formation mechanisms and mass evolution for extra-dimensional primordial black holes; the following two sections lay out a detailed calculation of greybody factors for both brane  (Sec.~\ref{sec:greybodybrane}) and bulk (Sec.~\ref{sec:greybodybulk}) degrees of freedom; finally, the results are used in Sec.~\ref{sec:predictions} to produce predictions for the stochastic background of high-frequency gravitational waves from brane-localized gravitons. We conclude in Sec.~\ref{sec:conclusions} with a summary and some remarks.

\section{Review of Large Extra Dimensions}

\subsection{Black Holes in Large Extra Dimensions}\label{sec:BHEDM}

The nature of black holes in extra-dimensional scenarios depends fundamentally on their size relative to the characteristic scale $R$ of the extra dimensions. Black holes with horizon radius $r_h>R$ are larger than the size of the extra dimensions, and so should be relatively insensitive\footnote{The presence of the extra dimensions still affects the evaporation spectrum of these black holes, as well as other properties.} to them, behaving effectively as 4-dimensional objects. In contrast, black holes with $r_h<R$ are fully $(n+4)$-dimensional objects. While primordial black holes can form through a variety of mechanisms in the early universe, we will primarily consider formation via energetic particle collisions in the hot primordial plasma. This mechanism is unique to the extra-dimensional scenario we will consider, since the lowering of the fundamental Planck scale implies that particle collisions with $E_{\rm CM} > M_*$ can lead to black hole formation. 

Once formed, the black hole proceeds through a series of phases as it evaporates \cite{Giddings:2001}. First during the \textit{balding phase}, the excited black hole will lose ``hair'' associated with gauge charges and multipole moments. The gauge charges, inherited from the initial state partons, are discharged via Schwinger emission. Quadrupole and higher multipole moments of the energy-momentum tensor associated with the initially asymmetric horizon are shed via classical gravitational radiation. This phase completes much more rapidly than the next two phases, which proceed via the emission of semi-classical Hawking radiation. The first is the \textit{spin-down phase}, wherein the black hole loses angular momentum and evolves towards a non-spinning spherically symmetric state. The black hole continues to lose energy during the final \textit{Schwarzschild phase}, leading to a gradual decrease in mass and increase in temperature. Since the majority\footnote{Although the majority of time is spent in the Schwarzschild phase, one should keep in mind that a decent portion of mass can be shed during the other phases. In 4-dimensions, the balding, spin-down, and Schwarzschild phases account for $\sim 16\%$, $25\%$, and $59\%$ of the mass loss, respectively \cite{Giddings:2001,Harris:2005}. (Note that these estimates can radically change in higher-dimensional spacetimes.) Since the balding and spin-down phases are dominated by graviton emission, the additional gravitational radiation from these phases can be even more significant. The results of this work should then be seen as a lower bound on graviton emission from black hole evaporation in extra-dimensional scenarios.} of evaporation is spent in the Schwarzschild phase \cite{Harris:2005}, this will be our focus in what follows.

A Schwarzschild black hole in $(n+4)$-dimensions is described by the line element \cite{Myers:1986}
\begin{equation}\label{4plusnDmetric}
    ds_{n+4}^2 = - h(r) dt^2 + h(r)^{-1} dr^2 + r^2 d\Omega_{2+n}^2 \,,
\end{equation}
where $d\Omega_{n+2}^2$ is the line element of the $(n+2)$-dimensional unit sphere and
\begin{equation}\label{hdef}
    h(r) = 1 - \left( \frac{r_h}{r} \right)^{n+1} \,.
\end{equation}
The horizon radius $r_h$ appearing in this expression can be written as a function of $n$, the black hole mass $M$, and $M_*$ as
\begin{equation}\label{radius}
    r_h = \frac{1}{M_*} \left( \frac{M}{M_*} \right)^{\frac{1}{n+1}} \left( \frac{8 \Gamma\big( (n+3)/2 \big)}{(n+2) \pi^{(n+1)/2}} \right)^{\frac{1}{n+1}} \,,
\end{equation}
and enters into the expression for the Hawking temperature associated with the black hole via
\begin{equation}\label{THdef}
    T_H=\frac{n+1}{4\pi r_h} \,.
\end{equation}
In the LED scenarios under consideration, black holes radiate gravitons into the bulk\footnote{The graviton zero mode is also confined to the brane.} and Standard Model particles on the brane. In the black body approximation, the power emitted in both forms scales as
\begin{equation}
    P \sim T_H^{n+4} r_h^{n+2} \sim T_H^2 \,,
\end{equation}
regardless of the number of dimensions. The black hole mass then evolves as $- \frac{dM}{dt} \sim T_H^2$, which can be solved for the lifetime
\begin{equation}
    \tau_{\rm BH} \sim \frac{1}{M_*} \left( \frac{M}{M_*} \right)^{\frac{n+3}{n+1}} \,.
\end{equation}
Comparing with the strictly 4-dimensional result, which scales as $\tau_{4\text{-dim}} \sim M^3/M_{\rm Pl}^4$, we see that an extra-dimensional black hole will usually\footnote{This is only strictly true for black holes satisfying $r_h<R$. When the radius becomes of order the size of the extra dimensions, one needs to carry out the full calculation including the numerical prefactors, which depend on $n$ and can be quite large.} be longer-lived than a 4-dimensional black hole of the same mass.

\subsection{Constraints on Extra-Dimensional Scenarios}\label{sec:constraints}

The constraints on the number of extra spatial dimensions and the bulk Planck scale $M_*$ are coupled and broadly originate from astrophysical and cosmological observations, precision tests of gravity at sub-mm distances, and collider searches. We summarize some of the constraints in Table \ref{tab:Mstar} and describe them below here. Note that scenarios with $n = 1$ are ruled out and those with $n=2$ are strongly constrained. However constraints are generally much weaker for $n > 2$ \cite{Workman:2022ynf}.

\begin{table}[]
    \centering
    \begin{tabular}{|c|c|c|}
    \hline
 $n$ & $M_* \geq$& Source\\
 \hline
         2&4 \unit{\TeV}& Tests of gravity  \\
          & 27 \unit{\TeV}& SN1987A\\
            & 1700 \unit{\TeV}& NS heating \\
         3&76  \unit{\TeV}&NS heating\\
         $\geq$ 4& $\sim 5$ \unit{\TeV}&Colliders \\
         \hline
         
    \end{tabular}
    \caption{A sample of constraints on the $(n+4)$-dimensional Planck scale $M_*$ for various numbers of large extra spatial dimensions $n$. See \cite{Workman:2022ynf} for a detailed review of the constraints.}
    \label{tab:Mstar}
\end{table}

Precision tests of gravity, i.e. deviations from Newton's law, limit the size of extra dimensions for $n=2$ at the level of $M_*>4.0$ TeV \cite{ref23}. Energy losses to Kaluza-Klein (KK) modes in stars, including supernovae, provide very stringent constraints: ref. ~\cite{ref25} calculates that energy losses to KK modes constrain $M_*>27$ TeV for $n=2$ and $M_*>2.4$ TeV for $n=3$. Since, after a supernova explosion, KK gravitons are gravitationally trapped in the remnant neutron star, heating from KK decays in the neutron stars constrains  $M_*>1700$ TeV for $n=2$ and $M_*>76$ TeV for $n=3$ \cite{ref25} (albeit decay to non-interacting ``dark sector'' degrees of freedom can weaken this bound). Cosmological constraints stem chiefly from the concern that particle production in the early universe (in particular, of relic KK gravitons) should not over-close the universe. This constrains, for instance, $M_*>7$ TeV for $n=2$ \cite{ref27}. Stronger bounds arise from distortion of the cosmic diffuse gamma radiation, however such constraints depend on whether KK gravitons can decay into dark sector states \cite{ref15}.

The most stringent collider constraints hinge on the production of KK graviton modes, or other modes that eventually decay into KK gravitons, which escape collider detection and manifest as missing transverse energy. While in the standard 4D case such processes are highly suppressed by the ``standard'' Planck scale, the reduced size of the true Planck scale $M_*$ makes them, in principle, testable at colliders such as the LHC. The constraints are increasingly weak with a larger number of extra spatial dimensions and result in an overall bound on the bulk Planck scale $M_*\gtrsim 5$ TeV for $n\ge 4$ \cite{Workman:2022ynf}. More specifically, the ATLAS analysis with 139 fb$^{-1}$ luminosity at a 13 TeV center-of-mass energy in \cite{ATLAS:2021kxv} quotes a 95\% C.L. limit of $M_*>11.2$ TeV for $n=2$ and of 
$M_*>5.9$ TeV for $n=6$. For a similar analysis with 137 fb$^{-1}$ luminosity, CMS quotes a 95\% C.L. limit of $M_*>10.7$ TeV for $n=2$ and of 
$M_*>5.5$ TeV for $n=6$ \cite{CMS:2021far}. 
\section{LED Black Hole Formation and Evaporation}\label{sec:edpbh} 

\subsection{Formation from High Energy Particle Collisions}\label{sec:collisions}

Primordial black holes can be formed through a variety of mechanisms in the early universe, including first order cosmological phase transitions, the collapse of topological defects, and the collapse of primordial overdensities (e.g. seeded by inflation). Regardless of the precise mechanism, it is suspected to be easier to produce black holes in extra-dimensional scenarios since $r_h^{(n+4)}>r_h^{(4)}$ means that matter on the brane requires less compression to form a horizon \cite{Argyres:1998}. In these extra-dimensional scenarios, tiny black holes can also be produced in high-energy particle collisions in the hot thermal plasma of the early universe, provided the center-of-mass energy of the colliding particles exceeds the quantum gravity scale $E_{\rm CM} \gtrsim M_*$. For a plasma of temperature $T \lesssim M_*$, black holes produced in this way have typical masses not much larger than the Planck scale $M \gtrsim M_*$, since the production of more massive black holes is Boltzmann suppressed. It was therefore initially thought that the tiny black holes formed through collisions would be too small and hot, evaporating essentially instantaneously and leaving no observational signatures. 

Refs.~\cite{Conley:2006,Friedlander:2022} and others have since noted that this need not be the case since accretion plays a vital role in the presence of extra dimensions. Since the accretion rate is proportional to the horizon area and since extra-dimensional black holes are generically larger than their 4-dimensional counterparts of the same mass, the effects of accretion cannot always be neglected. When accretion initially dominates over evaporation, these black holes can quickly grow to macroscopic size, asymptoting to a mass which is independent of the initial value. This asymptotic mass $M_{\rm asymp}$ depends on the nature of the bulk; for a bulk which is fully populated with gravitational radiation in thermal equilibrium with the brane, the mass will be that of a black hole with size of the extra dimensions \cite{Conley:2006}
\begin{equation}\label{MBHcollision1}
    M_{\rm asymp} = \frac{(n+2) \pi^{(n+1)/2}}{8 \Gamma[(n+3)/2]} \left( \frac{M_{\rm Pl}}{M_*} \right)^{2(n+1)/n} M_* \,.
\end{equation}
We will be more interested in the case that the bulk is not thermalized, in which case an initially microscopic ($r_h < R$) black hole will accrete matter at a rate \cite{Friedlander:2022}
\begin{equation}
    \frac{dM_{\rm acc}}{dt} = f_{\rm acc} \, A_h \, \rho_{\rm rad} \,,
\end{equation}
where $f_{\rm acc}$ is an $\mathcal{O}(1)$ accretion efficiency, $A_h = 4 \pi r_h^2$ is the horizon area, and $\rho_{\rm rad} = \frac{\pi^2}{30} g_{\star}(T) T^4$ is the radiation energy density, with $g_\star$ the number of relativistic degrees of freedom on the brane. It is convenient to re-express this as
\begin{equation}
    \frac{dM_{\rm acc}}{dt} = \left( \frac{\pi}{120} (n+1)^2 f_{\rm acc} g_\star \right) \left( \frac{T}{T_H} \right)^4 T_H^2 \equiv \chi_n \left( \frac{T}{T_H} \right)^4 T_H^2 \,.
\end{equation}
Meanwhile, black hole evaporation proceeds schematically at a rate
\begin{equation}
    \frac{dM_{\rm evap}}{dt} = - \alpha_n T_H^2 \,,
\end{equation}
where $\alpha_n$ encodes the greybody factors and depends only weakly on $T_H$ (see the following sections for details). The black hole mass evolves according to the sum of these terms
\begin{equation}
    \frac{dM}{dt} = \left[ - \alpha_n + \chi_n \left( \frac{T}{T_H} \right)^4 \right] T_H^2 \,.
\end{equation}
For initial black hole masses above the threshold $M_M^{\rm thresh}$, given as
\begin{equation}
    M_{\rm thresh} = \left( \frac{n+1}{4 \pi a_n} \left( \frac{\alpha_n}{\chi_n} \right)^{1/4} \frac{M_*}{T} \right)^{n+1} M_* \,,
\end{equation}
the latter term dominates the former, and the black hole preferentially accretes. In this case, we can approximate the evolution as $\frac{dM}{dt} \simeq \chi_n \frac{T^4}{T_H^2}$. During radiation domination
\begin{equation}
    H = \frac{1}{2t} = \left( \frac{4 \pi^3}{45} g_\star \right)^{1/2} \frac{T^2}{M_{\rm Pl}} \,,
\end{equation}
and so time and temperature can be related as
\begin{equation}
    \frac{dT}{dt} = - \left( \frac{4 \pi^3}{45} g_\star \right)^{1/2} \frac{T^3}{M_{\rm Pl}} \,.
\end{equation}
Keeping in mind also that $M$ appears in $T_H$ according to Eqs.~(\ref{radius}) and (\ref{THdef}), this differential equation can be solved for the asymptotic mass
\begin{equation}\label{MBHcollision2}
    M_{\rm asymp} = \left[ \left( \frac{\pi^3}{20} g_\star \right)^{1/2} \left( \frac{n-1}{n+1} \right) f_{\rm acc} a_n^2 \right]^{\frac{n+1}{n-1}} \left( \frac{M_{\rm Pl} T^2}{M_*^3} \right)^{\frac{n+1}{n-1}} M_* \,.
\end{equation}

While the initial fractional abundance $\Omega_{\rm BH}^i$ in the gravitational collapse formation scenario can essentially be treated as a free parameter, for collision-driven production it depends on the thermal distribution of particles in the plasma. Approximating the relative velocity between particles as relativistic, the black hole formation rate per unit mass is \cite{Conley:2006}
\begin{equation}\label{formationrate}
    \frac{dn_{\rm BH}}{dM} \simeq \frac{a_n^2 g_\star(T)^2}{8\pi^3} M T^2 \left( \frac{M}{M_*} \right)^{\frac{2(n+2)}{n+1}} \left[ \frac{M}{T} K_1 \left( \frac{M}{T} \right) + 2 K_2\left( \frac{M}{T} \right) \right] \Theta(M- M_*) \,,
\end{equation}
where $K_\nu(x)$ are modified Bessel functions of the second kind. Note that for low temperatures $T \ll M_* \lesssim M$, production becomes Boltzmann suppressed since $K_\nu\left(\frac{M}{T} \right) \simeq \sqrt{\frac{T}{M}} \exp\left( - \frac{M}{T} \right)$ in this limit. Thus the dominant production will occur at high temperatures, with the most efficient production at the reheating temperature $T_{\rm re}$. Despite the fact that this production would seem to give rise to an extended mass function, because rapid accretion leads to the asymptotic mass of either Eq.~(\ref{MBHcollision1}) in the case of a thermalized bulk or Eq.~(\ref{MBHcollision2}) otherwise, the result is a population of tiny primordial black holes with an essentially monochromatic mass function set by $n$, $M_*$, and the reheat temperature $T_{\rm re}$ alone. The initial fractional abundance $\Omega_{\rm BH}^i = \rho_{\rm BH}^i/\rho_{\rm tot}^i$ can be obtained from $\rho_{\rm BH} = M n_{\rm BH}$, with the number density obtained by integrating Eq.~(\ref{formationrate}) over $M$.

\subsection{Evaporation and Hawking Radiation}\label{sec:evaporation}

While accretion may initially dominate over evaporation for small black holes with $r_h < R$, once the temperature of the ambient plasma has fallen sufficiently relative to the Hawking temperature, evaporation will become the dominant process governing black hole evolution. A black hole with temperature $T_H$ evaporates by emitting Hawking radiation with a thermal spectrum which is almost that of a perfect blackbody. Explicitly for the Standard Model particles and zero-mode gravitons emitted on the brane, the flux spectrum is \cite{Hawking:1975}
\begin{equation}\label{fluxspectrum}
    \frac{dN^{(s)}(\omega)}{dt} = \sum_\ell \sigma_\ell^{(s)}(\omega) \frac{1}{e^{\omega/T_{\rm BH}} \mp 1} \frac{d^{3}k}{(2\pi)^{3}} \,,
\end{equation}
where the spin statistics factor is $\mp 1$ for bosons/fermions, respectively. Note that this expression applies for a given particle degree of freedom with spin $s$ and angular momentum $\ell$; to capture the contribution from an entire elementary particle, it is necessary to sum over the contributions from all spin or polarization degrees of freedom. The expression for the power spectrum, or energy emitted per unit time, is similarly
\begin{equation}\label{powerspectrum}
    \frac{dE^{(s)}(\omega)}{dt} = \sum_\ell \sigma_\ell^{(s)}(\omega) \frac{\omega}{e^{\omega/T_{\rm BH}} \mp 1} \frac{d^{3}k}{(2\pi)^{3}} \,.
\end{equation}
The degree to which this spectrum deviates from that of a perfect blackbody\footnote{For a perfect blackbody, the greybody factor is simply the area of the emitting body.} is parameterized by the greybody factor $\sigma_\ell^{(s)}(\omega)$. Intuitively, this factor appears because a particle emitted by the black hole needs to surmount the strong gravitational field about the black hole in order to reach an observer at infinity. The presence of this gravitational barrier distorts the shape of the spectrum from that of a blackbody distribution. It is conventional to relate the greybody factor to the absorption coefficient $\mathcal{A}_\ell^{(s)}(\omega)$ as
\begin{equation}\label{branegreybodydef}
    \sigma_\ell^{(s)}(\omega) = \frac{\pi}{\omega^2}\, \mathcal{N}_\ell \, \big| \mathcal{A}_\ell^{(s)}(\omega) \big|^2 \,,
\end{equation}
where $\mathcal{N}_\ell = 2\ell + 1$ is the multiplicity of states for partial wave $\ell$. The absorption coefficient in turn can be defined in terms of incoming and outgoing energy fluxes at the horizon $\mathcal{F}^H$ and at infinity $\mathcal{F}^\infty$,
\begin{equation}\label{absorption}
    \big| \mathcal{A}_\ell^{(s)}(\omega) \big|^2 = 1 - \frac{\mathcal{F}_{\rm out}^\infty}{\mathcal{F}_{\rm in}^\infty} = \frac{\mathcal{F}_{\rm in}^H}{\mathcal{F}_{\rm in}^\infty} \,,
\end{equation}
and thus can be computed by solving the equations of motion for particles propagating in this black hole background. This will be the task of the following sections.

Higher-dimensional black holes with horizons smaller than the characteristic scale of the extra dimensions $r_h < R$ also emit massive Kaluza-Klein scalar and graviton modes which propagate in the $(n+4)$-dimensional bulk. Even though only the Hawking radiation on the brane is visible, it is essential to account for emission in the bulk, as the energy lost to the bulk determines the energy remaining for brane emission. The expressions for particle and energy flux in this case are completely analogous
\begin{equation}\label{bulkfluxspectrum}
    \frac{d\hat{N}^{(s)}(\omega)}{dt} = \sum_\ell \hat{\sigma}_\ell^{(s)}(\omega) \frac{1}{e^{\omega/T_{\rm BH}} \mp 1} \frac{d^{n+3}k}{(2\pi)^{n+3}} \,,
\end{equation}
\begin{equation}\label{bulkpowerspectrum}
    \frac{d\hat{E}^{(s)}(\omega)}{dt} = \sum_\ell \hat{\sigma}_\ell^{(s)}(\omega) \frac{\omega}{e^{\omega/T_{\rm BH}} \mp 1} \frac{d^{n+3}k}{(2\pi)^{n+3}} \,,
\end{equation}
though note that we have introduced hats to signify bulk quantities. For bulk modes, the greybody factor is related to the absorption coefficient $\hat{\mathcal{A}}_\ell^{(s)}(\omega)$ as \cite{Gubser:1997}
\begin{equation}\label{bulkgreybodydef}
    \hat{\sigma}_\ell^{(s)}(\omega) = \frac{2^{n+1} \pi^{(n+1)/2} \Gamma \left( \frac{n+3}{2} \right)}{\omega^{n+2}} \, \hat{\mathcal{N}}_\ell \, \big| \hat{\mathcal{A}}_\ell^{(s)}(\omega) \big|^2 \,,
\end{equation}
where the combinatorial factor $\hat{\mathcal{N}}_\ell = \frac{(2\ell+n+1) (\ell+n)!}{\ell! (n+1)!}$ is the multiplicity of states for partial wave $\ell$ in $(n+4)$-dimensions. The definition of the absorption coefficient in terms of incoming and outgoing energy fluxes at the horizon and infinity is analogous to the 4-dimensional expression
\begin{equation}
    \big| \hat{\mathcal{A}}_\ell^{(s)}(\omega) \big|^2 = 1 - \frac{\hat{\mathcal{F}}_{\rm out}^\infty}{\hat{\mathcal{F}}_{\rm in}^\infty} = \frac{\hat{\mathcal{F}}_{\rm in}^H}{\hat{\mathcal{F}}_{\rm in}^\infty} \,.
\end{equation}

Though we are ultimately interested in the power spectrum of the zero-mode gravitons on the brane, which will give rise to the stochastic gravitational wave signal, the energy available to these gravitons will depend on that expended in emitting other brane-localized species as well as that lost to the bulk. Accurately modeling black hole evaporation thus requires a knowledge of greybody factors for all\footnote{More precisely, we will only consider particles of spin-0, 1/2, 1, and 2, since we restrict to the simplest realization of the LED scenario. In more involved scenarios, one may also need to consider spin-3/2 particles --- for example, the spin-3/2 gravitino in supersymmetric theories. While supersymmetry is not a requirement of LED scenarios, it is true that some of the best-motivated realizations of LED arise in supersymmetric theories, and so it would perhaps be prudent to also compute greybody factors for spin-3/2 particles. We leave this to future work.} species emitted. Determining these greybody factors is the goal of the following sections; we first calculate greybody factors for brane-localized modes in Sec.~\ref{sec:greybodybrane} before turning to bulk modes in Sec.~\ref{sec:greybodybulk}. The calculation is quite involved, so we summarize the main results in Table \ref{absorbcoefftable}. We also note that the greybody factors we compute match the results in the literature for scalars \cite{Kanti:2002}, fermions \cite{Kanti:2003}, and gauge bosons \cite{Kanti:2003} on the brane, as well as scalars \cite{Kanti:2002} and gravitons \cite{Creek:2006} in the bulk. The novel contribution of this work is that we also compute the greybody factor for the graviton zero-mode localized on the brane. We have verified that this expression reduces to the familiar 4-dimensional graviton greybody factor in the limit $n\rightarrow 0$ and $M_* \rightarrow M_{\rm Pl}$ by comparing the instantaneous graviton flux computed with our code in this limit with the output of the Hawking evaporation spectrum code \texttt{BlackHawk} \cite{Arbey:2019}.

\section{Greybody Factors I: Brane-Localized Modes}\label{sec:greybodybrane}

Brane-localized fields propagate in a gravitational background induced from the higher $(n+4)$-dimensional black hole solution, Eq.~(\ref{4plusnDmetric}). If we project this solution onto the brane by fixing the angular coordinates $\theta_i=\pi/2$ for the additional compact $n$ dimensions, the induced 4-dimensional line element takes the form
\begin{equation}\label{4Dmetric}
    ds_{4}^2 = - h(r) dt^2 + h(r)^{-1} dr^2 + r^2 d\Omega_{2}^2 \,, 
\end{equation}
with $h(r)$ given in Eq.~(\ref{hdef}). Greybody factors are obtained by solving master equations describing the motion of a particle of spin $s$ on this background. These master equations can be derived by making use of the Newman-Penrose formalism \cite{Newman:1962}, which is reviewed in Refs.~\cite{Kanti:2002,Kanti:2003,Kanti:2004}. 

We begin by factorizing the field as $\Psi_s = e^{-i\omega t} R_s (r) Y^s_{\ell m}(\theta)$, where the angular eigenfunctions $Y^s_{\ell m}(\theta) = e^{i m \varphi} S^s_{\ell m}(\theta)$ are spin-weighted spherical harmonics\footnote{The spin-weighted spherical harmonics $Y^s_{\ell m}(\theta)$ are generalizations of the usual spherical harmonics, and reduce to them for $s=0$.}. The radial equation for $R_s (r)$ reads \cite{Kanti:2003}
\begin{equation}\label{radeq1}
    \Delta^{-s} \frac{d}{dr} \left[ \Delta^{s+1} \frac{dR_s}{dr} \right] + \left[ \frac{\omega^2 r^2}{h} + i s \omega r \left( 2 - \frac{rh'}{h} \right) + s (\Delta'' - 2) - (2s-1)(s-1) r h' - \lambda_{s\ell} \right] R_s = 0 \,,
\end{equation}
where we define $\Delta = h r^2$ and $\lambda_{s\ell} = \ell(\ell + 1) - s(s+1)$ is a separation constant. This radial equation is almost identical to the usual 4-dimensional Teukolsky equation \cite{Teukolsky:1973}, but differs by the presence of the $s(\Delta''-2)$ term, which vanishes when $n=0$. 

We see that the radial equation and thereby also the greybody factors for brane-localized modes retain a dependence on the number of extra dimensions, which follows from the explicit appearance of $n$ in the line element induced by the higher-dimensional theory. Note also that this expression differs from the ``master equation'' quoted in Refs.~\cite{Cvetic:1997,Kanti:2004,Harris:2005} by the presence of the $(2s-1)(s-1) r h' R_s$ term. This vanishes for the spin-1/2 and spin-1 species considered in these works, but is non-vanishing for gravitons. While we will ultimately be interested in $s=2$ gravitons in the context of gravitational waves, we will also need the greybody factors for the other brane-localized species emitted by the black hole in order to faithfully model its evolution and evaporation. Thus we will leave $s$ arbitrary and solve Eq.~(\ref{radeq1}) in full generality.

This task will be easiest if we first make the change of variables $P_s = \Delta^s R_s$ to eliminate the $s \Delta''$ term, leading to the transformed radial equation
\begin{equation}\label{mastereq}
    \Delta^{s} \frac{d}{dr} \left[ \Delta^{1-s} \frac{dP_s}{dr} \right] + \left[ \frac{\omega^2 r^2}{h} + i s \omega r \left( 2 - \frac{rh'}{h} \right) - (2s-1)(s-1) r h' - \Lambda_{s\ell} \right] P_s = 0 \,,
\end{equation}
where for convenience we have also defined $\Lambda_{s\ell} = \lambda_{s \ell} + 2s = \ell(\ell + 1) - s(s-1)$. Despite this simplification, an exact analytic\footnote{Numerical integration of the radial master equation is also a possibility, but it is not without its own difficulties. In particular for higher spins, it becomes increasingly difficult to isolate the two asymptotic coefficients which enter into the expression for the greybody factor. Exact numerical solutions for $s=0$, $1/2$, and $1$ were derived in Ref.~\cite{Harris:2005}, however the numerical issues are such that no analysis has yet been performed for $s=2$.} solution is not possible for all $r$. It is possible, however, to solve in the near-horizon $r \simeq r_h$ and far-field $r \gg r_h$ regimes. This turns out to be sufficient for computing the greybody factors, which are related to the ratio of amplitudes for incoming and outgoing modes evaluated at either the horizon or infinity. The idea is to approximate the full solution of the radial equation by first solving in these limits, then stretching the solutions and matching in an intermediate regime.

More concretely, the asymptotic solution in the near-horizon limit $r \rightarrow r_h$ will turn out to take the form
\begin{equation}\label{NHform}
    P_s^H = A_{\rm in}^H e^{-i \omega r_*} + A_{\rm out}^H \Delta^{s} e^{i \omega r_*} \,,
\end{equation}
where $r_*$ is a tortoise coordinate satisfying $\frac{dr^*}{dr} = h^{-1}$. The condition that there be no outgoing modes near the black hole horizon requires us to set $A_{\rm out}^H = 0$. Meanwhile the asymptotic solution in the limit $r \rightarrow \infty$ will be of the form
\begin{equation}\label{FFform}
    P_s^{\infty} = A_{\rm in}^{\infty} \, \frac{e^{-i \omega r}}{(2 \omega r)^{1-2s}} + A_{\rm out}^{\infty} \, \frac{e^{i \omega r}}{2 \omega r} \,.
\end{equation}
In Secs.~\ref{sec:NH}, \ref{sec:FF}, and \ref{sec:match}, we solve the radial equation in the near-horizon and far-field regimes and then match the solutions in the intermediate region, which allows us to identify the coefficients $A_{\rm in}^H$, $A_{\rm in}^\infty$, and $A_{\rm out}^\infty$. Then in Sec.~\ref{sec:absorptioncoeff}, we use these coefficients to construct the absorption probabilities and ultimately the greybody factors. 

\subsection{Near-Horizon Regime}\label{sec:NH}

To solve Eq. (\ref{mastereq}) in the near-horizon regime, we begin by making the change of variables $r \rightarrow h$, in terms of which the equation becomes
\begin{equation}
\begin{split}
    & h(1-h) \frac{d^2 P_s}{dh^2} + \left[ (1-s)(1-h) - \frac{(n+2s)}{n+1} h \right] \frac{dP_s}{dh} +\\
    & \left[ \frac{\omega^2 r_h^2}{(n+1)^2 h (1-h)} + \frac{2 i s \omega r_h - \Lambda_{s \ell}}{(n+1)^2 (1-h)} - \frac{i s \omega r_h}{(n+1) h} - \frac{(2s-1)(s-1)}{n+1} \right] P_s = 0 \,.
\end{split}
\end{equation}
The further change of variables $P_s(h) = h^{\alpha} (1-h)^{\beta} F_s(h)$ transforms this into a hypergeometric equation of the generic form
\begin{equation}
    h(1-h) \frac{d^2 F_s}{dh^2} + \left[ c - (1 + a + b) h \right] \frac{dF_s}{dh} - a b F_s = 0 \,,
\end{equation}
with hypergeometric indices
\begin{subequations}\label{abcH}
\begin{equation}
    a = \alpha + \beta + \frac{1}{2(n+1)} \left[ s + n - ns + \sqrt{\kappa_{ns}} \right] \,,
\end{equation}
\begin{equation}
    b = \alpha + \beta + \frac{1}{2(n+1)} \left[ s + n - ns - \sqrt{\kappa_{ns}} \right] \,,
\end{equation}
\begin{equation}
    c = 1 - s + 2 \alpha \,,
\end{equation}
\end{subequations}
where $\kappa_{ns} = n^2 (s-1)^2 - 2n (5s-2)(s-1) - s(7s -12) - 4$. Substituting this Ansatz into Eq.~(\ref{mastereq}) and solving the resultant algebraic equations, one can identify the power coefficients
\begin{subequations}\label{alphabeta}
\begin{equation}
    \alpha_+ = s + \frac{i \omega r_h}{n+1} \,, \,\,\, \alpha_- = - \frac{i \omega r_h}{n+1} \,,
\end{equation}
\begin{equation}
    \beta_{\pm} = \frac{1}{2(n+1)} \left[ 1 - 2s \pm \sqrt{(2 \ell + 1)^2 - 4 \omega^2 r_h^2 - 8 i s \omega r_h} \right] \,.
\end{equation}
\end{subequations}
The full solution in the near-field regime reads
\begin{equation}
\begin{split}
    P_{\rm NH}^{(s)} = A_- h^{\alpha} (1 -h)^{\beta} F_s(a, b, c; h) + A_+ h^{- \alpha} (1-h)^{\beta} F_s(a - c + 1, b - c + 1, 2- c; h) \,,
\end{split}
\end{equation}
with $A_{\pm}$ as-yet-undetermined constants. In the limit $r \rightarrow r_h$, or equivalently $h \rightarrow 0$, this takes the asymptotic form
\begin{equation}
    \lim_{h \rightarrow 0} P_{\rm NH}^{(s)} \simeq A_- h^{\alpha} + A_+ h^{- \alpha} \,.
\end{equation}
Evaluating on the positive root $\alpha_+$, this becomes
\begin{equation}
    \lim_{h \rightarrow 0} P_{\rm NH}^{(s)} \big|_{\alpha_+} \simeq A_- h^s e^{i \omega r_*} + A_+ h^{- s} e^{- i \omega r_*} \,,
\end{equation}
where we have employed the tortoise coordinate $r_*$ introduced in Eq. (\ref{NHform}). This describes outgoing and incoming waves with divergent amplitudes at the horizon and so should be discarded. Meanwhile evaluating on the negative root $\alpha_-$ gives
\begin{equation}\label{Psh0}
    \lim_{h \rightarrow 0} P_{\rm NH}^{(s)} \big|_{\alpha_-} \simeq A_- e^{- i \omega r^*} + A_+ e^{i \omega r^*} \,,
\end{equation}
which is perfectly regular. Thus we choose $\alpha = \alpha_-$. Recall that we must also enforce the boundary condition that there be no outgoing modes at the horizon, instructing us to set $A_+ = 0$. Then comparing with the asymptotic form Eq. (\ref{NHform}), we see we should identify $A_{\rm in}^H \equiv A_-$. Finally, convergence of the hypergeometric function demands $\text{Re}[c - a - b] >0$, leading us to choose the negative root $\beta = \beta_-$. In summary, the solution in the near-field regime is
\begin{equation}\label{NHsol}
    P_{\rm NH}^{(s)} = A_{\rm in}^H h^{\alpha} (1 - h)^{\beta} F_s(a, b, c; h) \,,
\end{equation}
with hypergeometric indices $(a, b, c)$ as defined in Eq. (\ref{abcH}) and power coefficients $\alpha = \alpha_-$ and $\beta = \beta_-$ as defined in Eq. (\ref{alphabeta}). The amplitude for the incoming mode at the horizon $A_{\rm in}^H$ will be identified after matching the solutions in the intermediate regime, which requires first solving in the far-field limit.

\subsection{Far-Field Regime}\label{sec:FF}

Returning to Eq. (\ref{mastereq}) and now taking the far-field limit $r \gg r_h$, or equivalently $h \rightarrow 1$, the radial equation becomes
\begin{equation}
    \frac{d^2P_s}{dr^2} + \frac{2(1-s)}{r} \frac{dP_s}{dr} + \left[ \omega^2 + \frac{2 i s \omega}{r} - \frac{\Lambda_{s \ell}}{r^2} \right] P_s = 0 \,.
\end{equation}
Taking the Ansatz $P_s = e^{-i \omega r} r^{\ell + s} \tilde{P}_s$ and performing the change of variables $\rho = 2 i \omega r$, the equation adopts the confluent hypergeometric form
\begin{equation}
    \rho \frac{d^2 \tilde{P}_s}{d\rho^2} + (v - \rho) \frac{d \tilde{P}_s}{d \rho} - u \tilde{P}_s = 0 \,, 
\end{equation}
with indices
\begin{equation}
    u = \ell - s + 1 \,, \,\,\, v = 2(\ell + 1) \,.
\end{equation}
The general solution reads
\begin{equation}\label{FFsol}
    P_{\rm FF}^{(s)} = e^{-i \omega r} r^{\ell + s} \bigg( B_+ M(u, v; \rho) + B_- U(u, v; \rho) \bigg) \,,
\end{equation}
where $M$ and $U$ are Kummer functions, and we must keep both solutions. In order to fix the asymptotic normalization, consider taking the $r \rightarrow \infty$ limit, which leads to
\begin{equation}
\begin{split}
    \lim_{r \rightarrow \infty} P_{\rm FF}^{(s)} \simeq \Bigg[ \frac{e^{-i \pi(\ell - s +1)/2}}{(2\omega)^{\ell + s}} \bigg( B_- + & \frac{B_+ e^{i \pi(\ell - s + 1)} \Gamma[2\ell+2]}{\Gamma[\ell + s + 1]} \bigg) \Bigg] \frac{e^{-i \omega r}}{(2 \omega r)^{1-2s}}\\
    & + \left[ \frac{e^{-i \pi (\ell + s + 1)/2}}{(2 \omega)^{\ell + s}} \frac{B_+ \Gamma[2\ell+2]}{\Gamma[\ell - s + 1]} \right] \frac{e^{i \omega r}}{2 \omega r}\,,
\end{split}
\end{equation}
where the first and second terms are incoming and outgoing waves at infinity, respectively. Comparison with Eq. (\ref{FFform}) leads us to identify the bracketed quantity in the first line with $A_{\rm in}^\infty$,
\begin{equation}
    A_{\rm in}^\infty = \frac{e^{-i \pi(\ell - s +1)/2}}{(2\omega)^{\ell + s}} \bigg( B_- + \frac{e^{i \pi(\ell - s + 1)} \Gamma[2\ell+2]}{\Gamma[\ell + s + 1]} B_+ \bigg) \,,
\end{equation}
and that in the second line with $A_{\rm out}^\infty$,
\begin{equation}
    A_{\rm out}^\infty = \frac{e^{-i \pi (\ell + s + 1)/2}}{(2 \omega)^{\ell + s}} \frac{ \Gamma[2\ell+2]}{\Gamma[\ell - s + 1]} B_+ \,.
\end{equation}
Note that the incoming mode is dominant and the outgoing mode is suppressed as $1/r$ since this solution corresponds to the upper component\footnote{The other radiative component with $-s$ would have a suppressed incoming wave and dominant outgoing wave.} of the emitted field, with $+s$. 

\subsection{Intermediate Regime}\label{sec:match}

Finally, a complete solution and identification of the integration constants $A_-$ and $B_\pm$ requires us to match $P_{\rm NH}^{(s)}$ and $P_{\rm FF}^{(s)}$ in the intermediate region. Returning to the near horizon solution of Eq. (\ref{NHsol}), we want to stretch this towards large values of $r$. This will be easier if we first transform the argument of the hypergeometric function from $h$ to $1-h$ via the identity
\begin{equation}\label{hypergeometricidentity}
\begin{split}
    F(a,b,c;h) = & \frac{\Gamma(c) \Gamma(c-a-b)}{\Gamma(c-a) \Gamma(c-b)} F(a,b,1+a+b-c;1-h)\\
    & + \frac{\Gamma(c) \Gamma(a+b-c)}{\Gamma(a)\Gamma(b)} (1-h)^{c-a-b} F(c-a, c-b, 1+c-a-b; 1-h) \,.
\end{split}
\end{equation}
Now taking the limit $h \rightarrow 1$ ($r \rightarrow \infty$), the near-field solution becomes
\begin{equation}\label{NFstretchsol}
\begin{split}
    \lim_{h \rightarrow 1} P_{\rm NH}^{(s)} & \simeq A_{\rm in}^H \left( \frac{\Gamma[1-s+2\alpha] \Gamma[ \frac{1-2s}{n+1} -2 \beta]}{\Gamma[c-a] \Gamma[c-b]} \right) \left( \frac{r_h}{r} \right)^{(1-2s - \sqrt{(2\ell+1)^2 - 4 \omega^2 r_h^2 - 8 i s \omega r_h})/2}\\
    & + A_{\rm in}^H \left( \frac{\Gamma[1-s+2\alpha] \Gamma[2 \beta - \frac{1-2s}{n+1}]}{\Gamma[a] \Gamma[b]} \right) \left( \frac{r_h}{r} \right)^{(1-2s + \sqrt{(2\ell+1)^2 - 4 \omega^2 r_h^2 - 8 i s \omega r_h})/2} \,.
\end{split}
\end{equation}
As for the far-field solution of Eq. (\ref{FFsol}), stretching towards small values of $r \rightarrow r_h$ yields
\begin{equation}
    \lim_{r \rightarrow r_h} P_{\rm FF}^{(s)} \simeq B_+ r^{\ell + s} + \frac{B_-}{r^{\ell - s +1}} \frac{\Gamma[2\ell+1]}{(2i \omega)^{2\ell+1} \Gamma[\ell - s + 1]} \,.
\end{equation}
We would like to match this onto Eq. (\ref{NFstretchsol}), however these two expressions have different power law scalings in $r$. In order to eliminate the square root in the near-horizon solution, we must take the low energy limit $\omega r_h \ll 1$,
\begin{equation}
\begin{split}
    \lim_{h \rightarrow 1} P_{\rm NH}^{(s)} & \simeq A_{\rm in}^H \left( \frac{\Gamma[1-s+2\alpha]\, \Gamma[ \frac{1-2s}{n+1} -2 \beta]}{\Gamma[c-a]\, \Gamma[c-b]} \right) \left( \frac{r}{r_h} \right)^{\ell + s}\\
    & + A_{\rm in}^H \left( \frac{\Gamma[1-s+2\alpha] \, \Gamma[2 \beta - \frac{1-2s}{n+1}]}{\Gamma[a] \, \Gamma[b]} \right) \left( \frac{r_h}{r} \right)^{\ell - s + 1} \,.
\end{split}
\end{equation}
Note that this approximation is performed for the power law scalings only; no approximation is made to the gamma functions, which increases the domain of validity. Note also that we will ultimately be interested in the part of the spectrum ranging from the low frequency tail through the peak, which is set by the Hawking temperature $\omega_{\rm peak} \sim T_H$. From the fact that $T_H \sim 1/r_h$, we have that $\omega_{\rm peak} r_h \sim 1$, and so this approximation should roughly hold for our regime of interest. 

Finally we can identify
\begin{subequations}
\begin{equation}
    B_+ = \left( \frac{\Gamma[1-s+2\alpha]\, \Gamma[ \frac{1-2s}{n+1} -2 \beta]}{\Gamma[c-a] \,\Gamma[c-b]} \right) \frac{A_{\rm in}^H}{r_h^{\ell + s}} \,,
\end{equation}
\begin{equation}
    B_- = \left( \frac{\Gamma[1-s+2\alpha] \, \Gamma[2 \beta - \frac{1-2s}{n+1}]\, \Gamma[\ell - s + 1]}{\Gamma[a] \,\Gamma[b] \,\Gamma[2 \ell + 1]} \right) (2 i \omega r_h)^{2 \ell + 1} \frac{A_{\rm in}^H}{r_h^{\ell + s}} \,.
\end{equation}
\end{subequations}
Then the relation between incoming amplitudes at the horizon and infinity reads explicitly
\begin{equation}\label{infHratio}
\begin{split}
    A_{\rm in}^\infty = \frac{A_{\rm in}^H}{(2\omega r_h)^{\ell + s}} & \Bigg[ \frac{\Gamma(c)\, \Gamma\big(2\beta - \frac{1-2s}{n+1}\big)}{\Gamma(a)\, \Gamma(b)} \, \frac{(\ell -s)!}{(2\ell)!} \, (2 i \omega r_h)^{2\ell + 1} \, e^{-i \pi (\ell -s + 1)/2} \\
    & \hspace{19mm} + \frac{\Gamma(c)\, \Gamma\big( \frac{1-2s}{n+1} - 2 \beta\big)}{\Gamma(c-a)\, \Gamma(c-b)} \, \frac{(2\ell + 1)!}{(\ell + s )!} \, e^{i \pi (\ell -s + 1)/2} \Bigg] \,,
\end{split}
\end{equation}
with the hypergeometric indices $a,b,c$ defined in Eq.~(\ref{abcH}) and coefficients $\alpha,\beta$ in Eq.~(\ref{alphabeta}). 

\subsection{Absorption Coefficients}\label{sec:absorptioncoeff}

Having identified $A_{\rm in}^H$, $A_{\rm in}^\infty$, and $A_{\rm out}^\infty$, we are now ready to construct absorption coefficients $|\mathcal{A}_\ell^{(s)}|^2$, which were defined in Eq.~(\ref{absorption}) in terms of ratios of energy fluxes of incoming and outgoing modes at the horizon and infinity. In particular, we will use the second definition involving the incoming fluxes $\mathcal{F}_{\rm in}^\infty$ and $\mathcal{F}_{\rm in}^H$. This is more convenient since, for the upper field component we will consider, the outgoing mode at infinity is suppressed. 

For scalars and fermions, the incoming energy flux at the horizon or infinity can be computed by integrating the radial component of the conserved current $J^\mu$ over a 2-sphere at this location \cite{Cvetic:1997}. For a complex scalar $\Psi$, this conserved current is
\begin{equation}
    J^\mu_{(s=0)} = hr^2 (\Psi \partial^\mu \Psi^* - \Psi^* \partial^\mu \Psi) \,,
\end{equation}
leading to a flux proportional to the absolute square of the radial solution
\begin{equation}
    \mathcal{F}_{(s=0)} \sim \big|P_0 \big|^2 \,.
\end{equation}
Recall from Eq.~(\ref{NHform}) that the radial solution for an incoming mode at the horizon takes the generic form\footnote{This normalization is equivalent to that given in Eq.~(\ref{NHform}) upon solving for an explicit $r_*$ satisfying $dr_*/dr = h^{-1}$.}
\begin{equation}\label{NHincoming}
    P_s^H = A_{\rm in}^H h^{- i \omega r_h/(n+1)} \,,
\end{equation}
while from Eq.~(\ref{FFform}), the radial solution for an incoming mode at infinity is
\begin{equation}\label{FFincoming}
    P_s^\infty = A_{\rm in}^\infty \frac{e^{-i \omega r}}{(2 \omega r)^{1-2s}} \,.
\end{equation}
The absorption coefficient for a scalar is then
\begin{equation}\label{Ascalar}
    \big|\mathcal{A}_\ell^{(0)} \big|^2 = (2 \omega r_h)^2 \left| \frac{A_{\rm in}^H}{A_{\rm in}^\infty} \right|^2 \,.
\end{equation}
For a fermion $\Psi^A$ with conserved current
\begin{equation}
    J^\mu_{(s=1/2)} = \sqrt{2} \sigma_{AB}^\mu \Psi^A \bar{\Psi}^B \,,
\end{equation}
the flux is proportional to the difference of upper and lower field components 
\begin{equation}
    \mathcal{F}_{(s = 1/2)} \sim \big|P_{1/2} \big|^2 - \big|P_{-1/2} \big|^2 \,.
\end{equation}
Looking at Eq.~(\ref{FFincoming}), we see that for the lower field component, the incoming radial solution at infinity is suppressed. Thus the dominant contribution to the flux comes from the $s=+1/2$ helicity and the absorption coefficient is simply
\begin{equation}\label{Afermion}
    \big|\mathcal{A}_\ell^{(1/2)} \big|^2 = \left| \frac{A_{\rm in}^H}{A_{\rm in}^\infty} \right|^2 \,.
\end{equation}
For $s>1/2$, there are no conserved currents; however one can still infer the flow of energy from the energy momentum tensor. For a gauge boson $\Psi^{AB}$, one can use the $T^{rt}$ component of the energy momentum tensor
\begin{equation}
    T^{\mu \nu}_{(1)} = 2 \sigma_{AA'}^\mu \sigma_{BB'}^\nu \Psi^{AB} \bar{\Psi}^{A'B'} \,,
\end{equation}
to derive the flux
\begin{equation}
    \mathcal{F}_{s=1} \sim \frac{1}{2 \omega r^2} \left( \big|P_1 \big|^2 - \big| P_{-1} \big|^2 \right) \,. 
\end{equation}
Only the radial wavefunctions for $s=\pm 1$ appear because only the upper and lower field components are radiative. As with the $s=1/2$ case, the incoming mode is suppressed for the lower field component. Thus concentrating on the contribution from $s=+1$, the absorption coefficient is
\begin{equation}\label{Aboson}
    \big|\mathcal{A}_\ell^{(1)} \big|^2 = \frac{1}{(2\omega r_h)^2} \left| \frac{A_{\rm in}^H}{A_{\rm in}^\infty} \right|^2 \,.
\end{equation}
For scalars, fermions, and gauge bosons, the absorption coefficient can be conveniently summarized as
\begin{equation}\label{sabsorb}
    \big|\mathcal{A}_\ell^{(s)} \big|^2 = (2\omega r_h)^{2(1-2s)} \left| \frac{A_{\rm in}^H}{A_{\rm in}^\infty} \right|^2 \,,
\end{equation}
with the ratio $A_{\rm in}^H/A_{\rm in}^\infty$ given explicitly in Eq.~(\ref{infHratio}). Repeating this procedure for gravitons is complicated by the fact that there is generically no conserved energy-momentum tensor for spin-$2$. As advocated for in \cite{Cvetic:1997}, one option is to consider the Bel-Robinson tensor, from which one can infer an energy-momentum conserved in all static black hole spacetimes. In particular, the component $T^{rttt}$ can be used to construct the energy flux
\begin{equation}
    \frac{1}{4\pi} \frac{dE}{dt} = \frac{1}{16 \omega^2 r^6} \left( \big| P_2 \big|^2 - \big| P_{-2} \big|^2 \right) \,.
\end{equation}
Close to the horizon, we should replace the square of the frequency with $\omega_h^2 = \omega^2 + (2 \pi T_{\rm H})^2$, which comes from integrating with respect to proper distance along the worldline and is necessary to properly account for gravitons close to the horizon \cite{Cvetic:1997}. Again the negative helicity contribution is suppressed, so from the radial solution for $s=+2$, we have the absorption coefficient
\begin{equation}\label{gravitonabsorb}
    \big|\mathcal{A}_\ell^{(2)} \big|^2 = \frac{1}{64 \omega^4 \omega_h^2 r_h^6} \left| \frac{A_{\rm in}^H}{A_{\rm in}^\infty} \right|^2 \,,
\end{equation}
with $A_{\rm in}^H/A_{\rm in}^\infty$ in Eq.~(\ref{infHratio}). Substituting Eqs.~(\ref{sabsorb}) and (\ref{gravitonabsorb}) into Eq.~(\ref{branegreybodydef}) yields the greybody factors for all brane localized modes, and is the main result of this section.

\section{Greybody Factors II: Bulk Modes}\label{sec:greybodybulk}


\subsection{Bulk Scalars}

The calculation of greybody factors for bulk modes proceeds much in the same way as the calculation for the brane-localized modes, though the starting point is the complete $(n+4)$-dimensional metric of Eq.~(\ref{4plusnDmetric}). Since the scalar case is particularly straightforward, we will consider it first, following the treatment of \cite{Kanti:2002}. Let $\hat{\Psi}$ be a bulk scalar field, which can be factorized as $\hat{\Psi}(t,r,\Omega) = e^{-i \omega t} \hat{R}^{(0)}_{\omega \ell}(r) \hat{Y}_\ell(\Omega)$, where $\Omega = (\theta_i,\varphi)$ includes the angular coordinates and $\hat{Y}_\ell(\Omega)$ are $(n+3)$-dimensional spherical harmonics. The radial wavefunction satisfies the second-order differential equation
\begin{equation}\label{bulkradial}
    \frac{h}{r^{n+2}} \frac{d}{dr} \left( h r^{n+2} \frac{d\hat{R}^{(0)}}{dr} \right) + \left( \omega^2 - \frac{h}{r^2} \ell(\ell + n + 1) \right) \hat{R}^{(0)} = 0 \,,
\end{equation}
with $h(r)$ defined in Eq.~(\ref{hdef}). 

To solve in the near-horizon regime, we make the change of variables $r \rightarrow h$, such that the radial equation becomes
\begin{equation}
    h (1-h) \frac{d^2 \hat{R}^{(0)}}{dh^2} + (1-h) \frac{d\hat{R}^{(0)}}{dh} + \left( \frac{\omega^2 r^2}{(n+1)^2 h (1-h)} - \frac{\ell(\ell + n + 1)}{(n+1)^2 (1-h)} \right) \hat{R}^{(0)} = 0 \,.
\end{equation}
Near the horizon we may set $\omega^2 r^2 \rightarrow \omega^2 r_h^2$; then by taking the Ansatz $\hat{R}^{(0)} = h^{\hat{\alpha}} (1-h)^{\hat{\beta}} F(h)$, this takes the form of a hypergeometric equation
\begin{equation}
    h(1-h) \frac{d^2F}{dh^2} + [\hat{c} -(1 + \hat{a} + \hat{b})h] \frac{dF}{dh} - \hat{a} \hat{b} F = 0 \,,
\end{equation}
where $F(\hat{a}, \hat{b}, \hat{c}; h)$ is a hypergeometric function with indices $\hat{a} = \hat{b} = \hat{\alpha} + \hat{\beta}$ and $\hat{c} = 1 + 2 \hat{\alpha}$, and
\begin{subequations}
\begin{equation}
    \hat{\alpha}_\pm = \pm \frac{i \omega r_h}{n+1} \,,
\end{equation}
\begin{equation}
    \hat{\beta}_\pm = \frac{1}{2} \pm \frac{1}{2(n+1)} \sqrt{(2\ell + n + 1)^2 - (2 \omega r_h)^2} \,.
\end{equation}
\end{subequations}
Demanding convergence instructs us to choose the solution $\hat{\beta} = \hat{\beta}_-$ and as before we set $\hat{\alpha} = \hat{\alpha}_-$. The complete solution in the near-horizon regime is
\begin{equation}\label{bulkscalarNH}
    \hat{R}^{(0)}_{\rm NH} = \hat{A}_- h^{\hat{\alpha}} (1-h)^{\hat{\beta}} F(\hat{a}, \hat{b}, \hat{c}; h) + \hat{A}_+ h^{-\hat{\alpha}} (1-h)^{\hat{\beta}} F(\hat{a} -\hat{c} +1, \hat{b}-\hat{c}+1, 2 - \hat{c}; h) \,,
\end{equation}
and in the near-horizon limit $h \rightarrow 0$ this becomes
\begin{equation}
    \lim_{h \rightarrow 0} \hat{R}^{(0)}_{\rm NH} \simeq \left( \frac{r_h}{r} \right)^{(n+1) \hat{\beta}} \left( \hat{A}_- e^{- i \omega r_h^{n+2} y} + \hat{A}_+ e^{i \omega r_h^{n+2} y} \right) \,,
\end{equation}
with $y$ a tortoise coordinate satisfying $\frac{dy}{dr} = \frac{1}{h r^{n+2}}$. The boundary condition that there be no outgoing mode instructs us to set $\hat{A}_+ = 0$.

Returning to Eq.~(\ref{bulkradial}), to solve in the far-field regime $h \simeq 1$ we make the change of variables $\hat{R}^{(0)} = \hat{f}(r)/r^{(n+1)/2}$, in terms of which
\begin{equation}
    \left( \frac{d^2}{dr^2} + \frac{1}{r} \frac{d}{dr} + \left( \omega^2 - \frac{\ell (\ell+ n + 1)}{r^2} - \frac{(n+1)^2}{4 r^2} \right) \right) \hat{f} = 0 \,.
\end{equation}
This is a Bessel equation whose solution includes Bessel functions of the first $J_\nu(x)$ and second $Y_\nu(x)$ kind,
\begin{equation}\label{bulkscalarFF}
    \hat{R}^{(0)}_{\rm FF} = \frac{\hat{B}_+}{r^{(n+1)/2}} J_{\ell + (n+1)/2}(\omega r) + \frac{\hat{B}_-}{r^{(n+1)/2}} Y_{\ell + (n+1)/2}(\omega r) \,.
\end{equation}
In the asymptotic regime $\omega r \rightarrow \infty$, this takes the form
\begin{equation}\label{asympFF}
    \lim_{r \rightarrow \infty} \hat{R}^{(0)}_{\rm FF} \simeq \frac{\hat{B}_+ - i \hat{B}_-}{\sqrt{2\pi \omega r^{n+2}}} e^{i(\omega r - (2 \ell + n + 2)\pi/4)} + \frac{\hat{B}_+ + i \hat{B}_-}{\sqrt{2\pi \omega r^{n+2}}} e^{-i(\omega r - (2 \ell + n + 2)\pi/4)} \,.
\end{equation}

As before, the near-horizon solution Eq.~(\ref{bulkscalarNH}) needs to be stretched to $r \gg r_h$ and the far-field solution Eq.~(\ref{bulkscalarFF}) needs to be stretched to $r \ll \infty$ such that the solutions can be matched in the intermediate regime. Stretching the near-horizon solution can be achieved by first transforming the argument from $h$ to $1-h$ using the identity in Eq.~(\ref{hypergeometricidentity}) and then taking $h \rightarrow 1$, leading to
\begin{equation}
    \lim_{h\rightarrow 1} \hat{R}^{(0)}_{\rm NH} \simeq \hat{A}_- \left( \frac{\Gamma(1+2\hat{\alpha}) \, \Gamma(1 - 2 \hat{\beta})}{\Gamma(1 + \hat{\alpha} - \hat{\beta})^2} \right) \left( \frac{r}{r_h} \right)^{\ell} + \hat{A}_- \left( \frac{\Gamma(1+2\hat{\alpha}) \, \Gamma(2 \hat{\beta} - 1)}{\Gamma(\hat{\alpha} + \hat{\beta})^2} \right) \left( \frac{r_h}{r} \right)^{\ell + n + 1} \,.
\end{equation}
Stretching the far-field solution to smaller $r$ and taking the low energy limit $\omega r_h \ll 1$ again to match the powers of $r$ yields
\begin{equation}
    \lim_{r \rightarrow r_h} \hat{R}^{(0)}_{\rm FF} \simeq \hat{B}_+ \left( \frac{(\omega/2)^{\ell + (n+1)/2}}{\Gamma\big(\ell + \frac{n+3}{2} \big)} \right) r^\ell - \hat{B}_- \left( \frac{\Gamma\big(\ell + \frac{n+1}{2} \big)}{\pi (\omega/2)^{\ell + (n+1)/2}} \right) r^{-(\ell + n + 1)} \,.
\end{equation}
Matching the solutions then allows us to identify $\hat{B}_\pm$ in terms of $\hat{A}_-$. For convenience we define the ratio $\hat{B} \equiv \hat{B}_+/\hat{B}_-$, which is explicitly
\begin{equation}\label{Bhatdef}
    \hat{B} = - \frac{1}{\pi} \left( \frac{2}{\omega r_h} \right)^{2\ell + n + 1} \Gamma\big(\ell + \frac{n+1}{2}\big) \Gamma\big(\ell + \frac{n+3}{2} \big)\frac{\Gamma(\hat{\alpha}+\hat{\beta})^2}{\Gamma(1+ \hat{\alpha} - \hat{\beta})^2} \frac{\Gamma(1 - 2 \hat{\beta})}{\Gamma(2 \hat{\beta} - 1)} \,.
\end{equation}

We use the first expression\footnote{In the scalar case, both incoming and outgoing modes at infinity scale as $r^{-(n+2)/2}$, as can be seen from Eq.(\ref{asympFF}), and so neither is suppressed.} of Eq.~(\ref{absorption}) for the absorption coefficient, which depends on the ratio of incoming and outgoing fluxes at infinity
\begin{equation}
    \big| \hat{\mathcal{A}}_\ell^{(0)} \big|^2 = 1 - \frac{|\hat{R}^{(0)}_{\text{FF},\text{out}}|^2}{|\hat{R}^{(0)}_{\text{FF},\text{in}}|^2} \bigg|_\infty \,.
\end{equation}
In terms of the asymptotic far-field solutions of Eq.~(\ref{asympFF}), this is
\begin{equation}\label{absorbNS}
    \big| \hat{\mathcal{A}}_\ell^{(0)} \big|^2 = 1 - \bigg| \frac{\hat{B}_+ - i \hat{B}_-}{\hat{B}_+ + i \hat{B}_-} \bigg|^2 = \frac{4 \, \text{Im}[\hat{B}]}{|\hat{B}|^2 + 2 \, \text{Im}[\hat{B}]+1} \,.
\end{equation}

\subsection{Bulk Gravitons}
The bulk graviton perturbations should be decomposed into three parts: a symmetric traceless tensor piece, a vector piece, and a scalar piece. The radial equation obeyed by each type $(\hat{R}^T, \hat{R}^V, \hat{R}^S)$ is slightly different but can be solved in a manner exactly analogous to that used for the other perturbations we have considered so far. After making the change of variables to $h$ in order to solve in the near-horizon regime, the master equation \cite{Creek:2006} is 
\begin{equation}
    h(1-h) \frac{d^2 \hat{R}}{dh^2} + \left( 1- \frac{(2n+3)}{(n+1)} h \right) \frac{d\hat{R}}{dh} + \left( \frac{\omega^2 r_h^2}{(n+1)^2 h(1-h)} - \frac{\Lambda_G}{1-h} + C_G \right) \hat{R} = 0 \,, 
\end{equation}
where 
\begin{equation}
    \Lambda_G = \begin{cases} \frac{\ell(\ell + n + 1)}{(n+1)^2} + \frac{n(n+2)}{4(n+1)^2} & \hspace{5mm} \text{for} \,\,\, G=T,V \\ \frac{z}{16(n+1)^2 m^2} & \hspace{5mm} \text{for} \,\,\, G=S \end{cases} \,,
\end{equation}
and 
\begin{equation}
    C_G = \begin{cases} - \frac{(n+2)^2}{4(n+1)^2} & \hspace{5mm} \text{for} \,\,\, G=T \\ \frac{3 (n+2)^2}{4(n+1)^2} & \hspace{5mm} \text{for} \,\,\, G=V \\ - \frac{q(1-h)^2 + p(1-h) + w}{4(n+1)^2[2m + (n+2)(n+3)(1-h)]^2} & \hspace{5mm} \text{for} \,\,\, G=S \end{cases} \,,
\end{equation}
where for the scalar perturbations we have defined $m = \ell(\ell + n + 1) - n - 2$, $q = (n+2)^4 (n+3)^2$, $p = (n+2)(n+3) [4m(2n^2 + 5n + 6) + n(n+2)(n+3)(n-2)] - \frac{z(n+2)^2(n+3)^2}{4m^2}$, $w = -12m(n+2)[m(n-2) +n(n+2)(n+3)] - \frac{z(n+2)(n+3)}{m}$, and $z = 16m^3 + 4m^2(n+2)(n+4)$. For all perturbations, the solution obeying the boundary condition of no outgoing waves at the horizon is 
\begin{equation}
    \hat{R}^{T,V,S}_{\rm NH} = \hat{A}_- h^{\hat{\alpha}} (1-h)^{\hat{\beta}} F(\hat{a},\hat{b}, \hat{c}; h) \,,
\end{equation}
where
\begin{equation}
    \hat{\alpha} = - \frac{i \omega r_h}{n+1} \,, \,\,\, \hat{\beta} = \frac{1}{2(n+1)} \left( -1 - \sqrt{(2\ell + n + 1)^2 - 4 \omega^2 r_h^2} \right) \,,
\end{equation}
and the hypergeometric indices are
\begin{equation}
    \hat{a} = \hat{\alpha} + \hat{\beta} + \frac{n+2}{2(n+1)} + \lambda_G \,,\,\,\, \hat{b} = \hat{\alpha} + \hat{\beta} + \frac{n+2}{2(n+1)} - \lambda_G \,, \,\,\, \hat{c} = 1 + 2 \hat{\alpha} \,,
\end{equation}
with 
\begin{equation}
    \lambda_G = \begin{cases} 0 & \hspace{5mm} \text{for} \,\,\, G=T \\ \frac{(n+2)}{(n+1)} & \hspace{5mm} \text{for} \,\,\, G=V \\ \frac{(n+2)}{2(n+1)} \sqrt{1 - \frac{q + p + w}{[2m(n+2) + (n+2)^2(n+3)]^2}} & \hspace{5mm} \text{for} \,\,\, G=S \end{cases} \,.
\end{equation}
The far-field radial equation is much simpler, and for all perturbations takes the form
\begin{equation}
    \frac{d^2 \hat{R}^{T,V,S}}{dr^2} + \left( \omega^2 - \left( \ell(\ell + n + 1) + \frac{n(n+2)}{4} \right) \frac{1}{r^2} \right) \hat{R}^{T,V,S} = 0 \,,
\end{equation}
with solution
\begin{equation}
    \hat{R}_{\rm FF}^{T,V,S} = \hat{B}_+ \sqrt{r} J_{\ell + (n+1)/2}(\omega r) +\hat{B}_- \sqrt{r} Y_{\ell + (n+1)/2}(\omega r) \,.
\end{equation}
Stretching the solutions to the intermediate regime and matching allows for the identifications
\begin{equation}
    \frac{\hat{B}_+}{\hat{A}_-} = \left( \frac{2}{\omega r_h} \right)^{\ell + (n+1)/2} \frac{\Gamma\big( \ell + \frac{n+3}{2} \big) \Gamma(\hat{c}) \Gamma(\hat{c} - \hat{a} - \hat{b})}{\Gamma(\hat{c} - \hat{a}) \Gamma(\hat{c} - \hat{b}) r_h^{1/2}} \,,
\end{equation}
and
\begin{equation}
    \frac{\hat{B}_-}{\hat{A}_-} = - \pi \left( \frac{\omega r_h}{2} \right)^{\ell + (n+1)/2} \frac{\Gamma(\hat{c}) \Gamma(\hat{a} + \hat{b} - \hat{c})}{\Gamma\big( \ell + \frac{n+1}{2} \big) \Gamma(\hat{a}) \Gamma(\hat{b}) r_h^{1/2}} \,.
\end{equation}
Again, we define the ratio $\hat{B} = \hat{B}_+/\hat{B}_-$, so
\begin{equation}\label{hatBdef2}
    \hat{B} = -\frac{1}{\pi} \left( \frac{2}{\omega r_h} \right)^{2\ell + n + 1} \frac{\Gamma \left( \ell + \frac{n+3}{2} \right) \Gamma \left( \ell + \frac{n+1}{2} \right) \Gamma(\hat{a}) \Gamma(\hat{b}) \Gamma(\hat{c} - \hat{a} - \hat{b})}{\Gamma(\hat{c} - \hat{a}) \Gamma (\hat{c} - \hat{b}) \Gamma(\hat{a} + \hat{b} - \hat{c}) } \,.
\end{equation}
For the absorption coefficient, we expand the far-field solution in the asymptotic limit $r \rightarrow \infty$ and use the definition in terms of incoming and outgoing fluxes at infinity. The solution is the same as in the bulk scalar case,
\begin{equation}
    \big| \hat{\mathcal{A}}_\ell^{T,V,S} \big|^2 = 1 - \bigg| \frac{\hat{B}_+ - i \hat{B}_-}{\hat{B}_+ + i \hat{B}_-} \bigg|^2 = \frac{4 \, \text{Im}[\hat{B}]}{|\hat{B}|^2 + 2 \, \text{Im}[\hat{B}]+1} \,,
\end{equation}
but now with $\hat{B}$ given in Eq.~(\ref{hatBdef2}).

\begin{table}
\centering
\begin{tabular}{|c||c|c|}
 \hline \xrowht[()]{14pt}
 \textbf{Particle species} & \textbf{Absorption coefficient} $\mathbf{\big|\phat{\mathcal{A}}_\ell \big|^2}$ & \textbf{Relevant equations} \\ 
 \hline\hline \xrowht[()]{12pt}
 Scalar, $s=0$ & $(2 \omega r_h)^2 \left| A_{\rm in}^H/A_{\rm in}^\infty \right|^2$ & \multirow{6}{*}{with $\left| A_{\rm in}^H/A_{\rm in}^\infty \right|^2$ in Eq.~(\ref{infHratio})} \\ 
 \cline{1-2} \xrowht[()]{12pt}
 Fermion, $s=1/2$ & $\left| A_{\rm in}^H/A_{\rm in}^\infty \right|^2$ &  \\
 \cline{1-2} \xrowht[()]{12pt}
 Gauge boson, $s=1$ & $\frac{1}{(2\omega r_h)^2} \left| A_{\rm in}^H/A_{\rm in}^\infty \right|^2$ & \\
 \cline{1-2} \xrowht[()]{12pt}
 Zero-mode graviton, $s=2$ & $\frac{1}{64 \omega^4 \omega_h^2 r_h^6} \left| A_{\rm in}^H/A_{\rm in}^\infty \right|^2$ & \\
 \hline\hline \xrowht[()]{12pt}
 Bulk scalar & \multirow{6}{*}{\text{\Large $\frac{4 \, \text{Im}[\hat{B}]}{|\hat{B}|^2 + 2 \, \text{Im}[\hat{B}]+1}$}} & with $\hat{B}$ in Eq.~(\ref{Bhatdef}) \\ 
 \cline{1-1}\cline{3-3} \xrowht[()]{12pt}
 Graviton, scalar perturbation &  & \multirow{4}{*}{with $\hat{B}$ in Eq.~(\ref{hatBdef2})} \\
 \cline{1-1} \xrowht[()]{12pt}
 Graviton, vector perturbation &  &  \\
 \cline{1-1} \xrowht[()]{12pt}
 Graviton, tensor perturbation &  &  \\
 \hline
\end{tabular}
\caption{Summary of absorption coefficients for degrees of freedom on the brane and in the bulk.}
\label{absorbcoefftable}
\end{table}

\section{Emission Rates and Gravitational Wave Spectra}\label{sec:predictions}

The goal of this paper is to study the emission of brane-localized (zero-mode) gravitons during the Hawking evaporation of primordial black holes in large extra dimensions. Ultimately we would like to compute the gravitational wave spectra from this source, but since the energy available to brane-localized gravitons depends on that expended in emitting all other species, it is necessary to consider the emission rates for these as well. 

\begin{figure}[h!]
\centering
\includegraphics[width=0.48\textwidth]{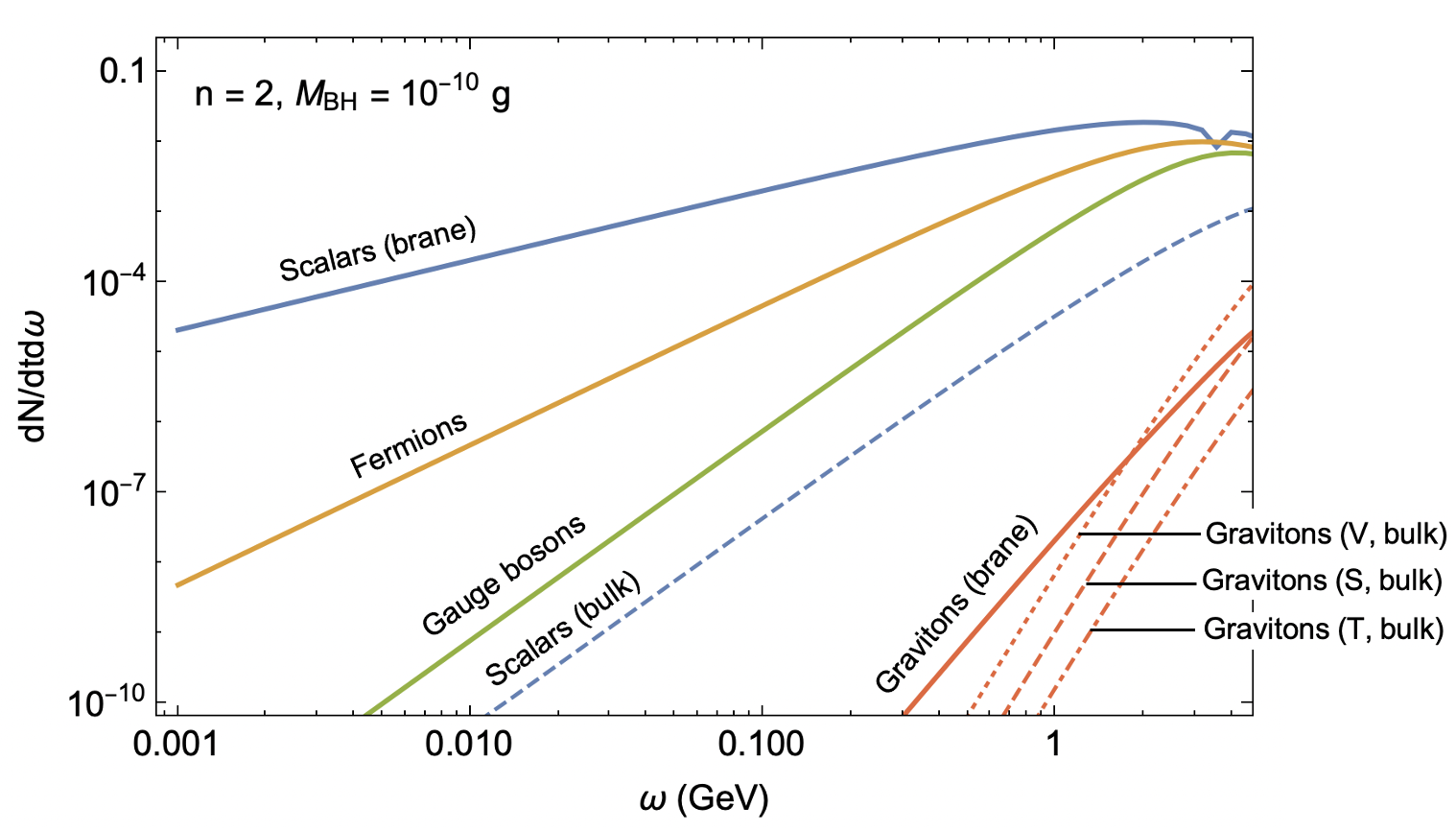}
\hspace{2mm}
\includegraphics[width=0.48
\textwidth]{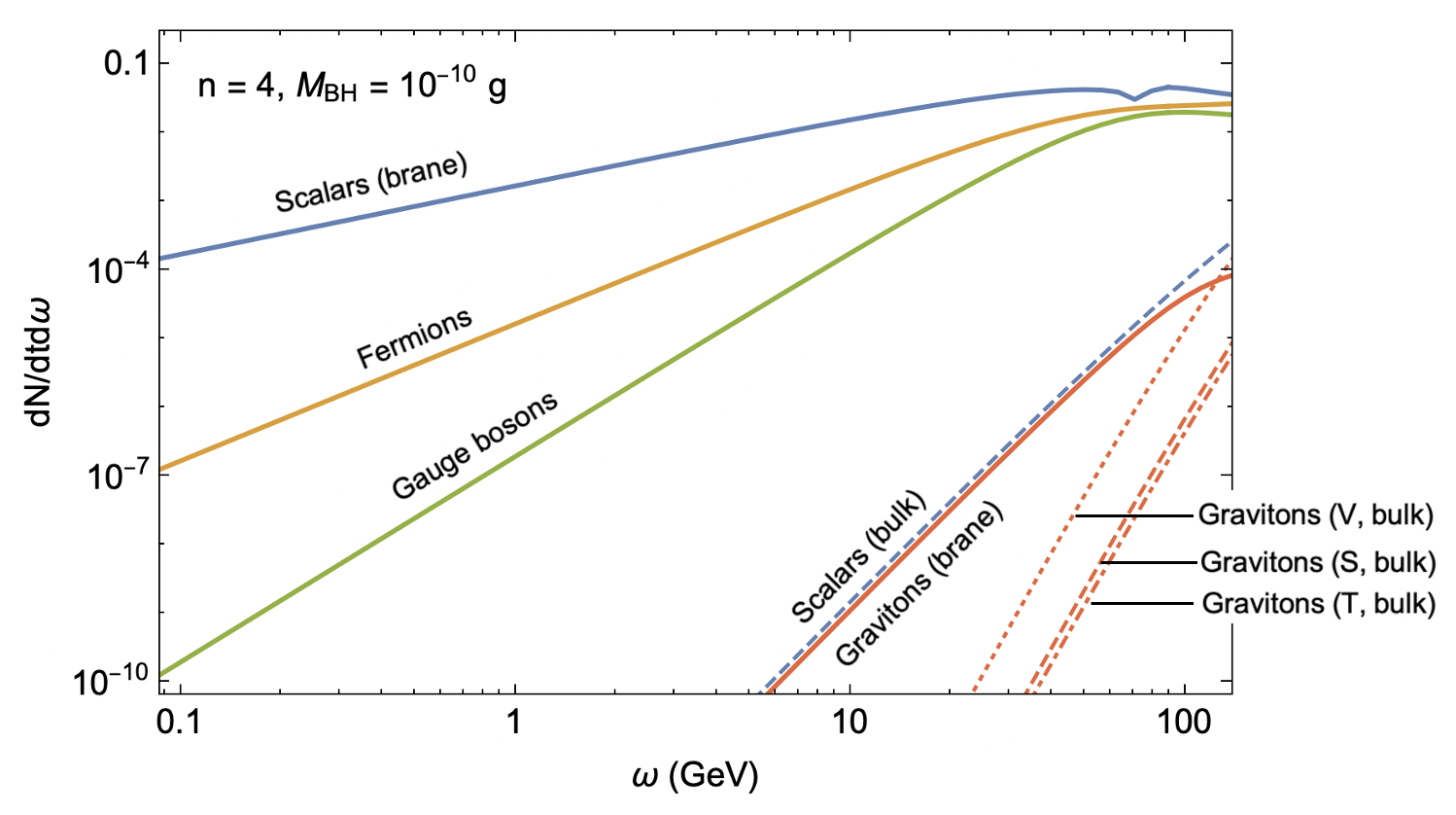}
\caption{Instantaneous flux $\frac{dN}{dt d\omega}$ of species on the brane and in the bulk for a representative black hole of mass $M = 10^{-10} \, \text{g}$, bulk Planck scale $M_\star = 10\, \text{TeV}$, and $n=2$ and $n=4$ extra dimensions, respectively. The same qualitative trends are observed for the power spectra $\frac{dE}{dt d\omega}$. The right-hand side of the $x$-axis is truncated where the low frequency approximation breaks down, $\omega r_h \sim 1$. The dent in the brane scalar line at high-energies is a numerical feature coming from the break down of this approximation.}
\label{fluxcomparison}
\end{figure}

The emission of particles with mass greater than the Hawking temperature is exponentially suppressed, so for convenience, we consider only the emission of species\footnote{The mass of the lightest KK scalar is set by the size of the extra dimensions as $m_{KK} \simeq 1/R$, which in turn is set by our choice of $M_*$ and $n$ since $M_{\rm Pl}^2 \sim R^n M_*^{n+2}$. In particular for $M_* = 10^3$ TeV and $n=4$, we expect the lightest mode to have GeV-scale mass.} with $m<T_{\rm BH}$ and treat these as massless, $|\vec{k}| = \omega$. This way, the phase space integrals in the flux and power spectra of Eqs.~(\ref{fluxspectrum})-(\ref{powerspectrum}) and Eqs.~(\ref{bulkfluxspectrum})-(\ref{bulkpowerspectrum}) can be evaluated explicitly. Using also Eqs.~(\ref{branegreybodydef}) and (\ref{bulkgreybodydef}) to re-express the greybody factors in terms of absorption coefficients, the flux and power spectra simplify considerably to
\begin{subequations}
\begin{equation}
    \frac{dN^{(s)}}{dt d\omega} = \frac{1}{2\pi} \sum_\ell (2\ell + 1) \big| \mathcal{A}_\ell^{(s)} \big|^2 \frac{1}{e^{\omega/T_{\rm BH}} \mp 1} \,, 
\end{equation}
\begin{equation}\label{dEdtdomega}
    \frac{dE^{(s)}}{dt d\omega} = \frac{1}{2\pi} \sum_\ell (2\ell + 1) \big| \mathcal{A}_\ell^{(s)} \big|^2 \frac{\omega}{e^{\omega/T_{\rm BH}} \mp 1} \,, 
\end{equation}
\end{subequations}
on the brane and 
\begin{subequations}
\begin{equation}
    \frac{d\hat{N}^{(s)}}{dt d\omega} = \frac{1}{2\pi} \sum_\ell \frac{(2\ell+n+1)(\ell + n)!}{\ell!(n+1)!} \, \big| \hat{\mathcal{A}}_\ell^{(s)} \big|^2 \frac{1}{e^{\omega/T_{\rm BH}} \mp 1} \,, 
\end{equation}
\begin{equation}
    \frac{d\hat{E}^{(s)}}{dt d\omega} = \frac{1}{2\pi} \sum_\ell \frac{(2\ell+n+1)(\ell + n)!}{\ell!(n+1)!} \, \big| \hat{\mathcal{A}}_\ell^{(s)} \big|^2 \frac{\omega}{e^{\omega/T_{\rm BH}} \mp 1} \,, 
\end{equation}
\end{subequations}
in the bulk. Fig.~\ref{fluxcomparison} below compares emission rates for the various species for two sample benchmark points.
We see that at higher frequencies, the differences in emission rate for all brane localized species become smaller, as one would expect since in the high frequency limit, the greybody factors for all species approach the geometric optics limit. Note that we truncate the $x$-axis at the frequency at which the low frequency approximation used in deriving the greybody factors breaks down. At low frequencies, the emission of particles with higher spin is suppressed due to the larger barrier such particles have to surmount.

In general, the emission rate of species on the brane exceeds that of bulk species, and this effect becomes more pronounced with increasing number of extra dimensions $n$. This is perhaps surprising, since both black hole temperature and the multiplicity of states are enhanced at higher $n$. However the absorption probability is suppressed with increasing $n$, and ultimately it is this effect that dominates. Thus for all bulk degrees of freedom, we observe a low energy emission rate which decreases with increasing $n$, consistent with the findings of \cite{Creek:2006}. We also confirm these authors' claim that among the gravitational perturbations in the bulk, vector-type dominate.


Now we turn to calculate the gravitational wave signal from this source, as parameterized by the spectral density parameter $\Omega_{\rm GW}$, defined as
\begin{equation}
    \Omega_{\rm GW} = \frac{1}{\rho_{\rm crit}} \frac{d\rho_{\rm GW}}{d \ln f} \,.
\end{equation}
Our starting point is the instantaneous power spectrum for a single degree of freedom for a brane-localized graviton, $\frac{dE^{(2)}}{dt d\omega}$, the general expression for which is given in Eq.~(\ref{dEdtdomega}) with $|\mathcal{A}_\ell^{(2)}|^2 = \frac{1}{64 \omega^4 \omega_h^2 r_h^6}|A_{\rm in}^H/A_{\rm in}^\infty|^2$ and  $|A_{\rm in}^H/A_{\rm in}^\infty|^2$ in Eq.~(\ref{infHratio}). Multiplying by 2 to account for the 2 graviton polarizations we define $\frac{dE_{\rm GW}}{dt d\omega} = 2 \frac{dE^{(2)}}{dt d\omega}$. For an entire population of black holes with number density $n_{\rm BH}$, the instantaneous energy density emitted in gravitational waves is then
\begin{equation}
    \frac{d\rho_{\rm GW}}{dt d\omega} = n_{\rm BH} \frac{dE_{\rm GW}}{dt d\omega} \,.
\end{equation}
To obtain the total energy density in zero-mode gravitons emitted over the black hole lifetime, we integrate this expression from black hole formation at $t_i$ to evaporation at $t_* = t_i + \tau_{\rm BH}$, where the black hole lifetime $\tau_{\rm BH}$ can be obtained by numerically following the black hole evolution 
\begin{equation}
    \frac{dM}{dt} = - \sum_{\text{all\,d.o.f.}} \int d\omega \left( \frac{d\phat{E} \text{\tiny \textcolor{white}{l}}^{(s)}}{dt d\omega} \right) \,,
\end{equation}
from $M = M_i$ to $\sim M_*$. For practical purposes when integrating $\frac{d\rho_{\rm GW}}{dt d\omega}$, it is useful to make the time dependence of certain quantities explicit. In particular recall that $\rho_{\rm GW} \sim a^{-4}$, $\omega \sim a^{-1}$, and $n_{\rm BH} \sim a^{-3}$. The integrated energy density at evaporation is then
\begin{equation}
    \frac{d\rho_{\rm GW}^*}{d \ln \omega} = n_{\rm BH}^i \omega_* a_i^3 \int_{t_i}^{t_*} dt \, a(t)^{-3} \left( \frac{dE_{\rm GW}}{dt d\omega} \right) \,,
\end{equation}
Note that we have also converted the frequency interval to a logarithmic frequency interval. We presume that black hole formation occurs during radiation domination and define $\Omega_{\rm BH}^i$ as the initial fractional energy density in black holes, in terms of which the initial number density can be expressed
\begin{equation}
    n_{\rm BH}^i = \frac{3M_{\rm Pl}^2 \Omega_{\rm BH}^i}{32 \pi M_i t_i^2} \,.
\end{equation}
We also allow for the possibility that the black holes come to dominate at some time $t_{\rm eq}$, which is approximately
\begin{equation}
    t_{\rm eq} \simeq \left( \frac{1-\Omega_{\rm BH}^i}{\Omega_{\rm BH}^i} \right)^2 t_i \,.
\end{equation}
The scale factor appearing in this expression then scales with time as
\begin{equation}
    a(t) = \begin{cases}
        a_i \left( \frac{t}{t_i} \right)^{1/2} & t < t_{\rm eq} \\ a_i \left( \frac{t_{\rm eq}}{t_i} \right)^{1/2} \left( \frac{t}{t_{\rm eq}} \right)^{2/3} & t> t_{\rm eq} \,.
    \end{cases}
\end{equation}
Finally to obtain the gravitational wave spectrum today, we need to account for the redshift in energy density and frequency due to the cosmological expansion between evaporation and today
\begin{equation}
    \frac{d\rho_{\rm GW}^0}{d \ln f} = \frac{d\rho_{\rm GW}^*}{d \ln \omega} \left( \frac{a_*}{a_0} \right)^4 \,.
\end{equation}
We take the scale factor today to be $a_0=1$, which can be related to that at evaporation $a_*$ by invoking the conservation of entropy $g_{\star,s} a^3 T^3 = \text{constant}$,
\begin{equation}
    a_* = \left( \frac{g_{\star,s}(T_0)}{g_{\star,s}(T_*)} \right)^{1/3} \frac{T_0}{T_*} \,,
\end{equation}
where $T_0 = 0.235 \,\text{meV}$, $g_{\star,s}(T_0) = 3.91$, and $T_*$ is the temperature when evaporation concludes. In the event that black holes dominate, this can be estimated by equating the energy density in black holes right before decay with that in radiation immediately afterwards, leading to 
\begin{equation}
    T_* \simeq \left( \frac{5 M_{\rm Pl}^2}{\pi^3 g_\star(T_*) t_*^2} \right)^{1/4} \,.
\end{equation}
The energy density in gravitational waves today per logarithmic frequency interval is then
\begin{equation}
    \frac{d \rho_{\rm GW}^0}{d \ln f} = \frac{g_{\star,s}(T_0)}{g_{\star,s}(T_*)} \left( \frac{T_0}{T_*} \right)^3 \omega_0 n_{\rm BH}^i a_i^3 \int_{t_i}^{t_*} dt \, a(t)^{-3} \left( \frac{dE_{\rm GW}}{dt d\omega} \right) \,,
\end{equation}
and the spectral density parameter is
\begin{equation}
    \Omega_{\rm GW} = \frac{\Omega_{\rm BH}^i \omega_0^4}{H_0^2 M^i} \mathcal{I}(\omega_0) \,,
\end{equation}
where $H_0 = 100 h \text{km}\cdot \text{s}^{-1} \cdot \text{Mpc}^{-1}$ is the Hubble rate and we have pulled out the leading frequency scaling, defining the integral
\begin{equation}
    \mathcal{I}(\omega_0) = \frac{1}{4 t_i^2} \frac{g_{\star,s}(T_0)}{g_{\star,s}(T_*)} \left( \frac{T_0}{T_*} \right)^3 \frac{1}{\omega_0^3} \int_{t_i}^{t_*} dt \, \left( \frac{a_i}{a(t)} \right)^3 \left( \frac{dE_{\rm GW}}{dt d\omega} \right) \,.
\end{equation}
Finally for the sake of comparing against projected sensitivities, we will sometimes present our results in terms of the dimensionless characteristic stain $h_c$, which is related to the spectral density parameter as
\begin{equation}
    \Omega_{\rm GW} = \frac{4\pi^2}{3 H_0^2} f^2 h_c^2 \,.
\end{equation}

\begin{figure}[h!]
\centering
\includegraphics[width=0.75\textwidth]{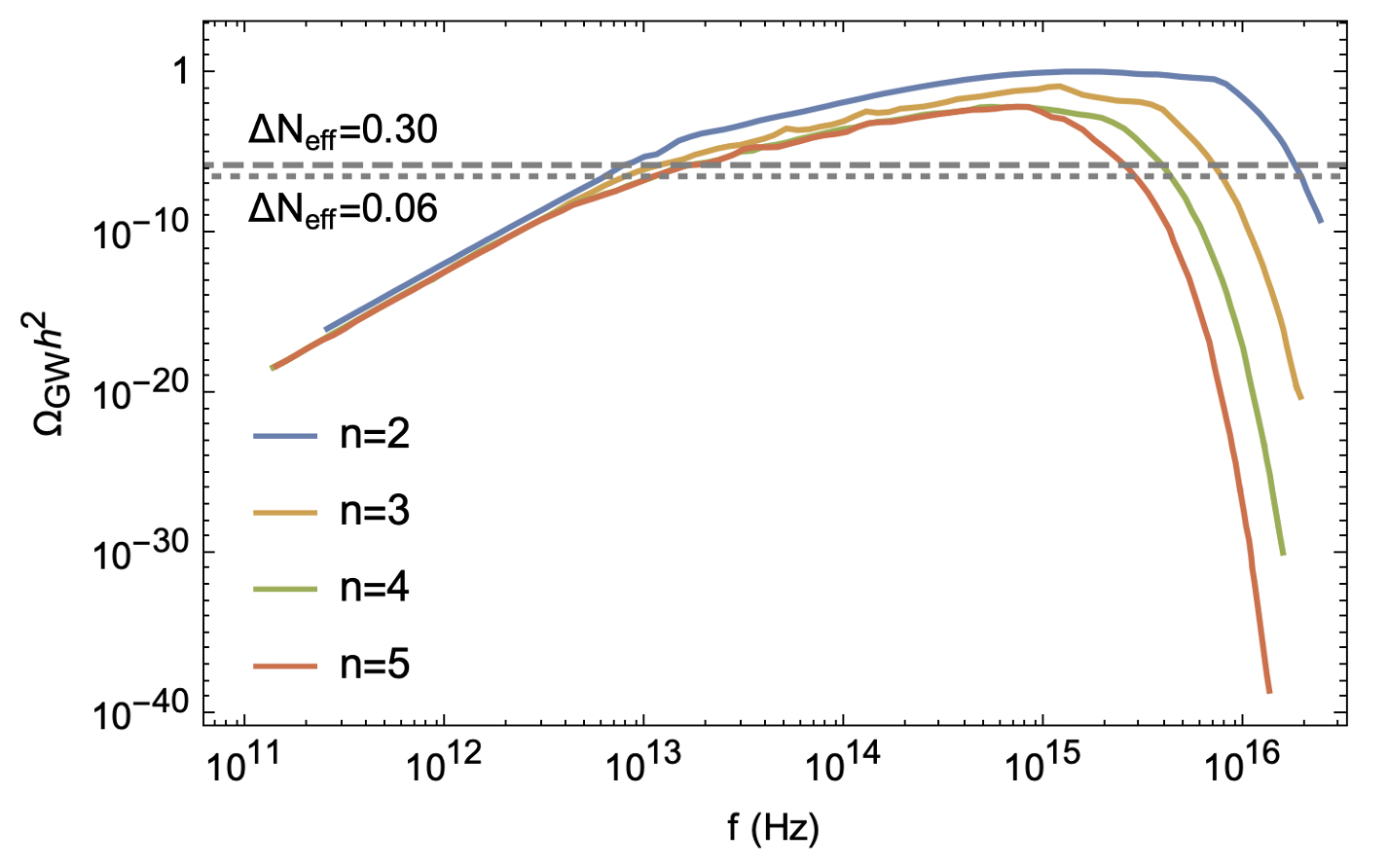}
\caption{\noindent Gravitational wave spectra (in terms of the spectral density parameter $\Omega_{\rm GW}h^2$) for various numbers of extra dimensions $n=2,3,4,5$ and a benchmark set of parameters: $M_*=10^3\, \text{TeV}$, $t_i=10^{-30}\, {\rm s}$, $M=1 \, \text{g}$, $T_{\rm re}=10^{5}\,\text{GeV}$.}
\label{fig:changingN}
\end{figure}

In Fig.~\ref{fig:changingN}, we plot predictions for the gravitational wave spectrum for a benchmark point with bulk Planck scale $M_*=10^3\, \text{TeV}$, formation time $t_i=10^{-30}\, {\rm s}$, initial black hole mass $M=1 \, \text{g}$, reheating temperature $T_{\rm re}=10^{5}\,\text{GeV}$, and various numbers of extra spatial dimensions $n$.
We indicate by the dashed horizontal line the integral bound on the total energy density in gravitational waves. The observable $\Delta N_{\rm eff}$ parameterizes the additional amount of radiation energy density beyond that of photons. In particular the contribution from gravitational waves reads
\begin{equation}
    \Delta N_{\rm eff} = \frac{8}{7} \left( \frac{11}{4} \right)^{4/3} \frac{\rho_{\rm GW}}{\rho_\gamma} \,,
\end{equation}
and so the \textit{Planck} constraint of $\Delta N_{\rm eff} < 0.30$ \cite{Planck:2018} can be used to bound the maximum amplitude of the gravitational wave signal $\Omega_{\rm GW} < 3.6\times 10^{-6}$ \cite{Ireland:2023}. We indicate by the dotted horizontal line the projected sensitivity from the future CMB-stage 4 experiment, $\Delta N_{\rm eff} < 0.06$ at 95\% CL \cite{CMBS4:2022}, which would constrain $\Omega_{\rm GW} < 7.2\times 10^{-7}$. Finally, we normalize the initial\footnote{Note that here we do not restrict to formation via high-energy particle collisions, in which case $\Omega_{\rm BH}^i$ would be completely fixed by this choice of parameters. Instead, since we are interested in obtaining an upper bound on the signal, we consider all production channels, in which case we can treat $\Omega_{\rm BH}^i$ as a free parameter.} energy density in black holes $\Omega_{\rm BH}^i$ such that prior to decay, the black holes come to dominate the energy density of the universe. Thus these plots should be interpreted as giving an \textit{upper bound} on the maximum possible signal. As the number of extra dimensions is increased, the spectrum is qualitatively similar but increasingly suppressed at high frequencies. 

\begin{figure}[h!]
\centering
\includegraphics[width=1.1\textwidth]{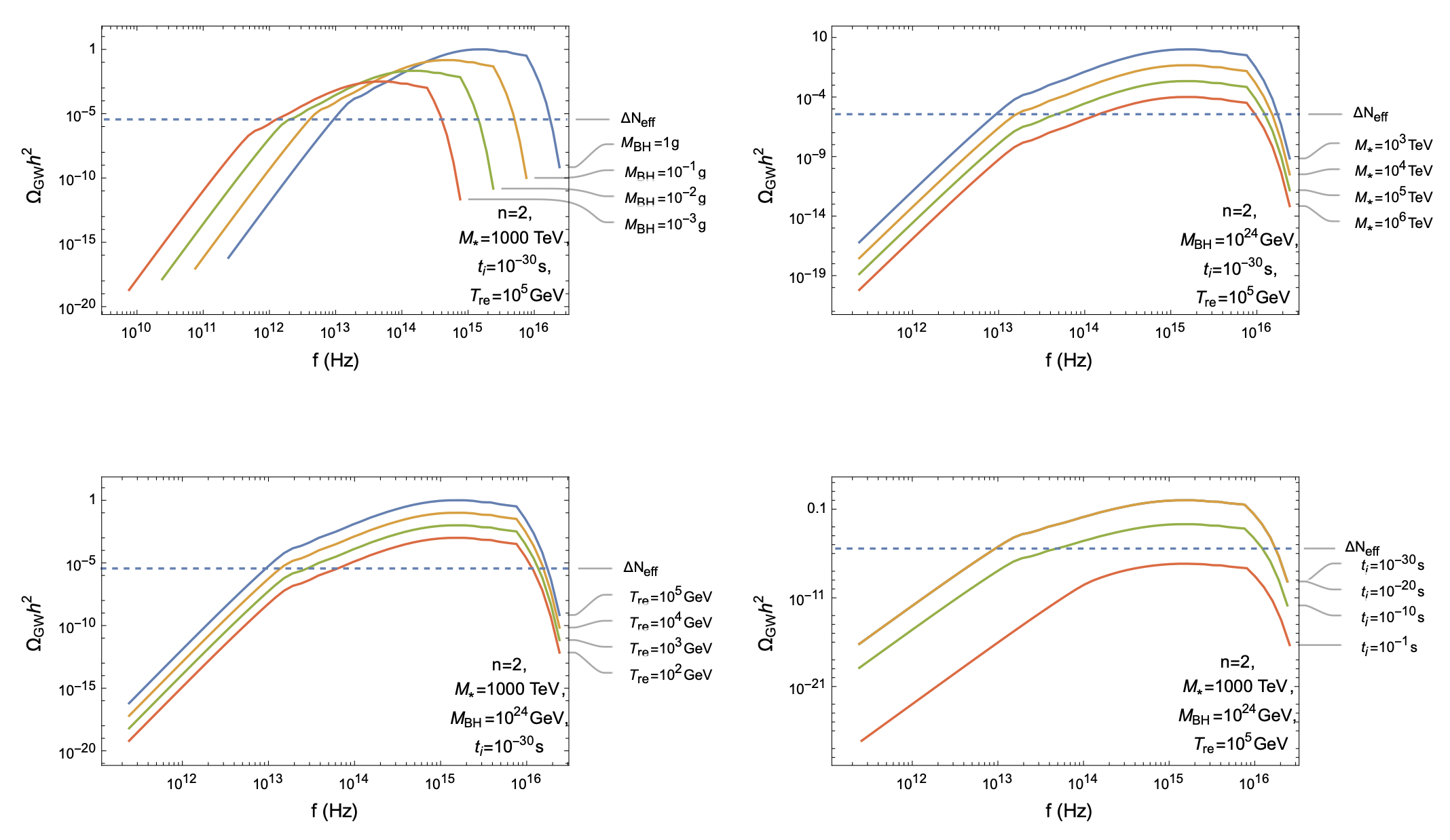}
\caption{\noindent Gravitational wave spectra (in terms of the spectral density parameter $\Omega_{\rm GW}h^2$) for $n=2$ and the base benchmark set of Fig.~\ref{fig:changingN}. In each panel we vary a single parameter. Top left: varying $M$; Top right: varying $M_*$; Bottom left: varying $T_{\rm re}$; Bottom right: varying $t_i$. Note that in the bottom right panel, the blue line corresponding to $t_i = 10^{-30}\,\text{s}$ is essentially coincident with the yellow $t_i = 10^{-20}\,\text{s}$.}
\label{fig:changingparameters}
\end{figure}

In Fig.~\ref{fig:changingparameters}, we fix $n=2$ and illustrate the effect of changing other parameters.
In the top left panel, we vary $M$; in the top right, $M_*$; in the bottom left, $T_{\rm re}$; and in the bottom right, $t_i$. A few comments are in order. First, decreasing the black hole mass moves the spectrum towards lower frequencies, which makes sense as smaller black holes evaporate more promptly, leading to a longer period of cosmological redshifting of the signal. This also serves to suppress the maximum amplitude of the signal. Next, we see that decreasing the Planck scale $M_*$ enhances the overall signal while leaving the location of the peak frequency unaffected. A similar effect is observed upon either increasing the reheating temperature $T_{\rm re}$ or decreasing the formation time $t_i$. 

\begin{figure}[h!]
\centering
\includegraphics[width=1\textwidth]{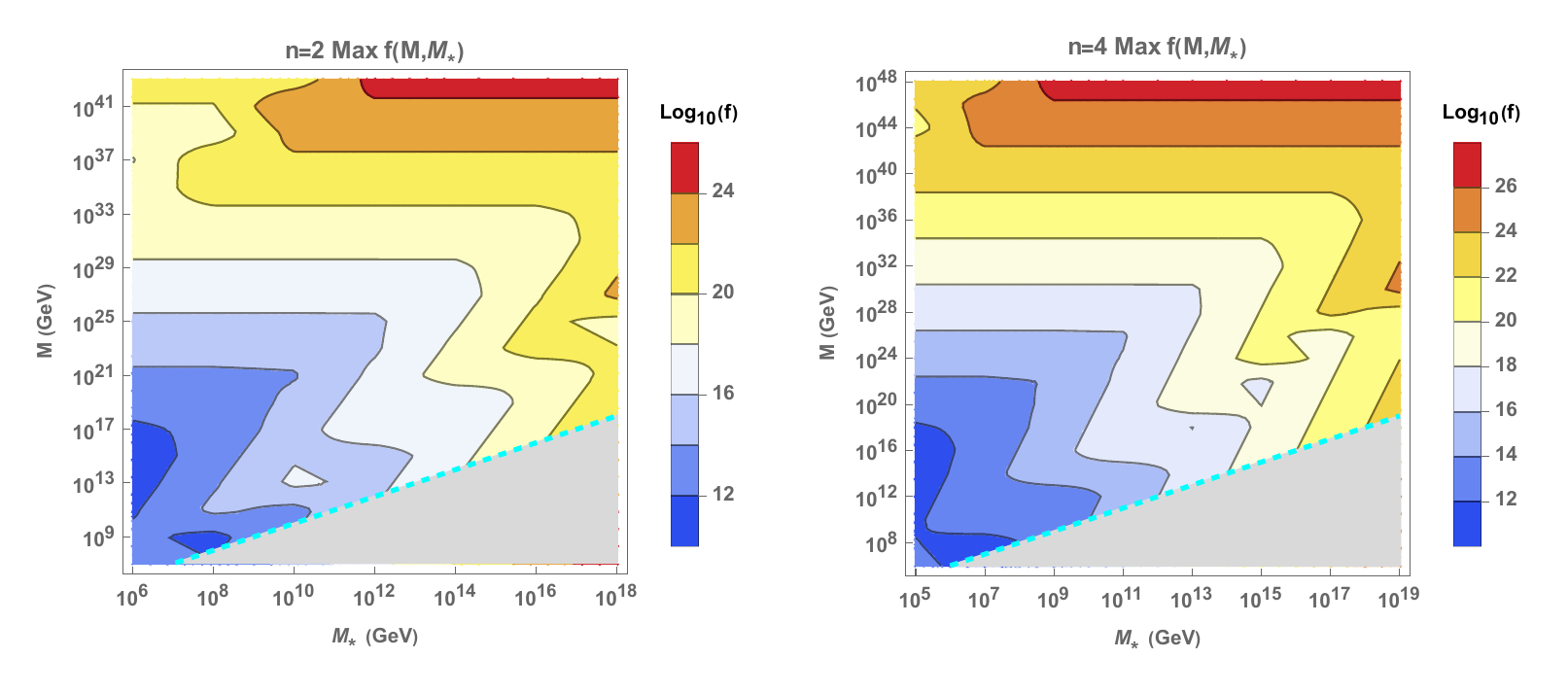}
\caption{Contours of constant peak frequency in the plane defined by the reduced Planck scale $M_*$ and the black hole mass $M$. Warmer colors indicate higher peak frequencies. Left panel: $n=2$; Right panel: $n=4$. We set the other parameters to the benchmark values of Fig.~\ref{fig:changingN}. The grey region is excluded on the basis that $M<M_*$.}
\label{fig:frequencycontours}
\end{figure}

\begin{figure}[h!]
\centering
\includegraphics[width=0.9\textwidth]{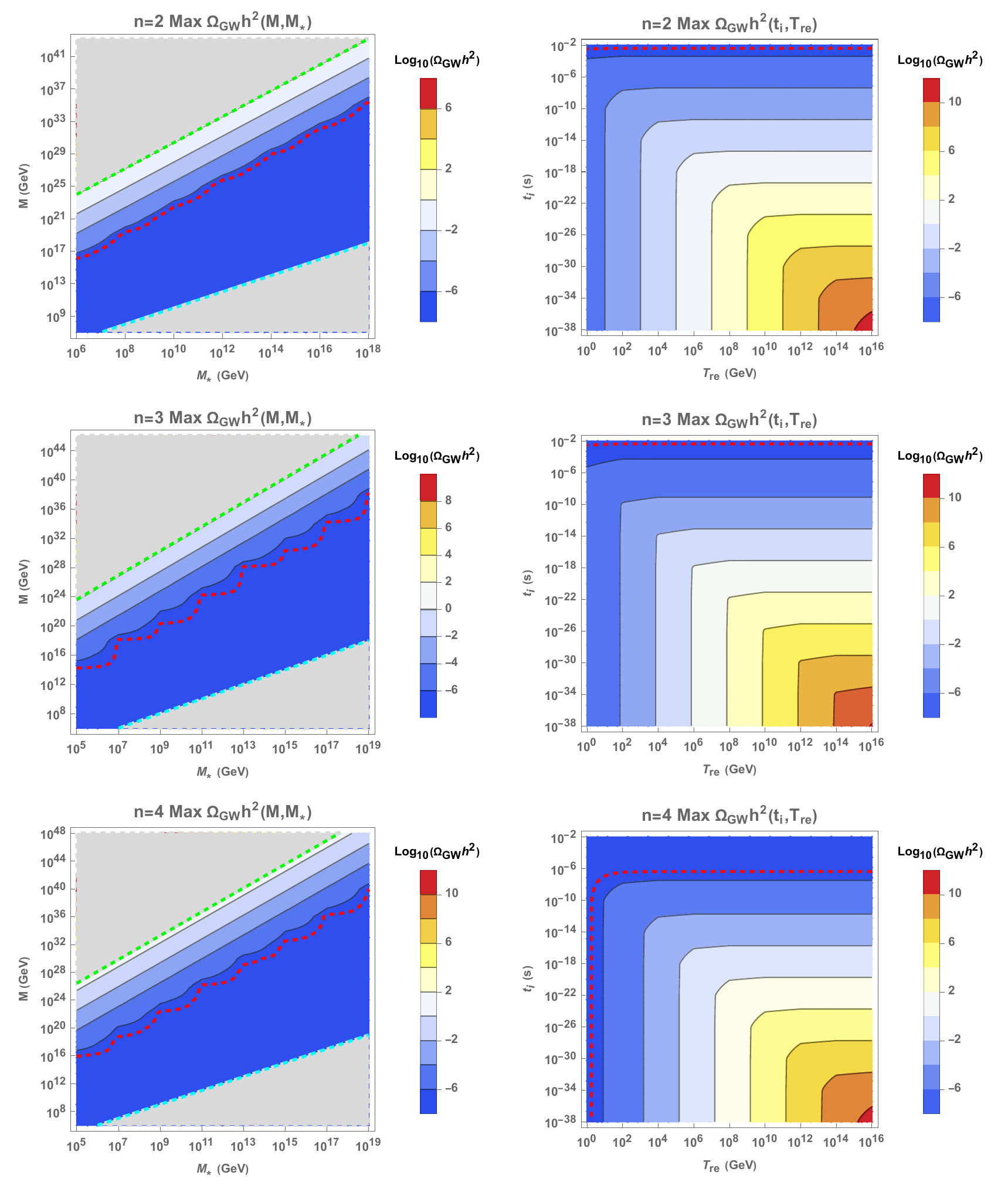}
\caption{Contours of constant gravitational wave energy density $\Omega_{\rm GW}h^2$ at peak frequency in the ($M_*,M$)-plane (left panels) at fixed $t_i=10^{-30}\, {\rm s}$ and $T_{\rm re}=10^{5}\,\text{GeV}$, and in the ($T_{\rm re},t_i$)-plane (right panels) at fixed $M_*=10^3 \,\text{TeV}$ and $M=1 \,\text{g}$. As in the previous figure, warmer colors indicate larger amplitudes. The top panels have $n=2$, the middle panels $n=3$, and the bottom panels $n=4$ extra dimensions. Amplitudes corresponding to the constraint on $\Delta N_{\rm eff}$ are shown with red dashed lines, with larger amplitudes excluded; the green dashed lines indicate parameter space where evaporation occurs during BBN, with larger amplitudes ruled out; finally, the cyan dashed lines indicate ``quantum black holes'' (i.e. those with $M=M_*$), with the grey area below excluded.}
\label{fig:amplitudecontours}
\end{figure}

We want to identify the ``optimal'' scenario for experimental detection --- i.e. the lowest frequency signal with maximal amplitude consistent with $\Delta N_{\rm eff}$ constraints. Clearly there are quite a few parameters that can be varied, so to determine the overall trends, in Figs.~\ref{fig:frequencycontours} and \ref{fig:amplitudecontours} we present contour plots of the peak frequency $f$ and maximal spectral density parameter $\Omega_{\rm GW} h^2$, respectively, for various slicings of parameter space. In Fig.~\ref{fig:changingparameters} we saw that changing the formation time and reheating temperature had minimal effect on the value of the peak frequency, which instead was primarily set by the choice of the reduced Planck scale $M_*$ and the black hole mass $M$. Thus in Fig.~\ref{fig:frequencycontours} we set $t_i=10^{-30}\, {\rm s}$ and $T_{\rm re}=10^{5}\, \text{GeV}$ and explore how peak frequency varies in the $(M_*,M)$-plane for $n=2$ (left) and $n=4$ (right). For both cases, the lowest values of the peak frequency correspond to low values of both $M_*$ and $M$ (bottom left corners). Increasing either\footnote{It may appear surprising that increasing $M$ leads to higher frequency, since from Eq.~(\ref{THdef}) we have that the Hawking temperature peaks at $T_H \sim (M_*/M)^{1/(n+1)}M_*$. This only sets the peak frequency at emission, though, and after redshifting the signal to today, it turns out that the peak frequency roughly scales as $f_{\rm peak}^0 \sim T_0 (M/M_{\rm Pl})^{1/2}$, which is indeed increasing with increasing $M$. It is perhaps surprising that $M_*$ does not appear in this expression, but this estimate was derived under many approximations (instant decay, blackbody emission), and it is expected that $M_*$-dependence enters at subleading order. Looking at the frequency plots, the $M_*$-dependence indeed appears quite weak, substantiating this reasoning.} leads to peak gravitational wave emission at higher frequency.

Fig.~\ref{fig:amplitudecontours} shows contour plots of the maximum gravitational wave amplitude in the $(M_*,M)$- and $(T_{\rm re}, t_i)$-planes for $n=2$ (top), $n=3$ (middle), and $n=4$ (bottom) extra dimensions. 
The dashed green line and the grey region above it correspond to pairs of $(M_*,M)$ that lead to evaporation at times larger than $1\, \text{s}$, violating BBN constraints. As in Fig.~\ref{fig:frequencycontours}, the cyan dashed line and the grey region below should be excluded on the basis that $M<M_*$. The red dashed line corresponds to the $\Delta N_{\rm eff}$ bound, with larger amplitudes excluded. We note that this rules out a very significant portion of the otherwise permissible parameter space, including nearly all $t_i$ for $n=2$ and $n=3$ at the benchmark $M$ and $M_*$.

\begin{figure}[h!]
\centering
\includegraphics[width=0.8\textwidth]{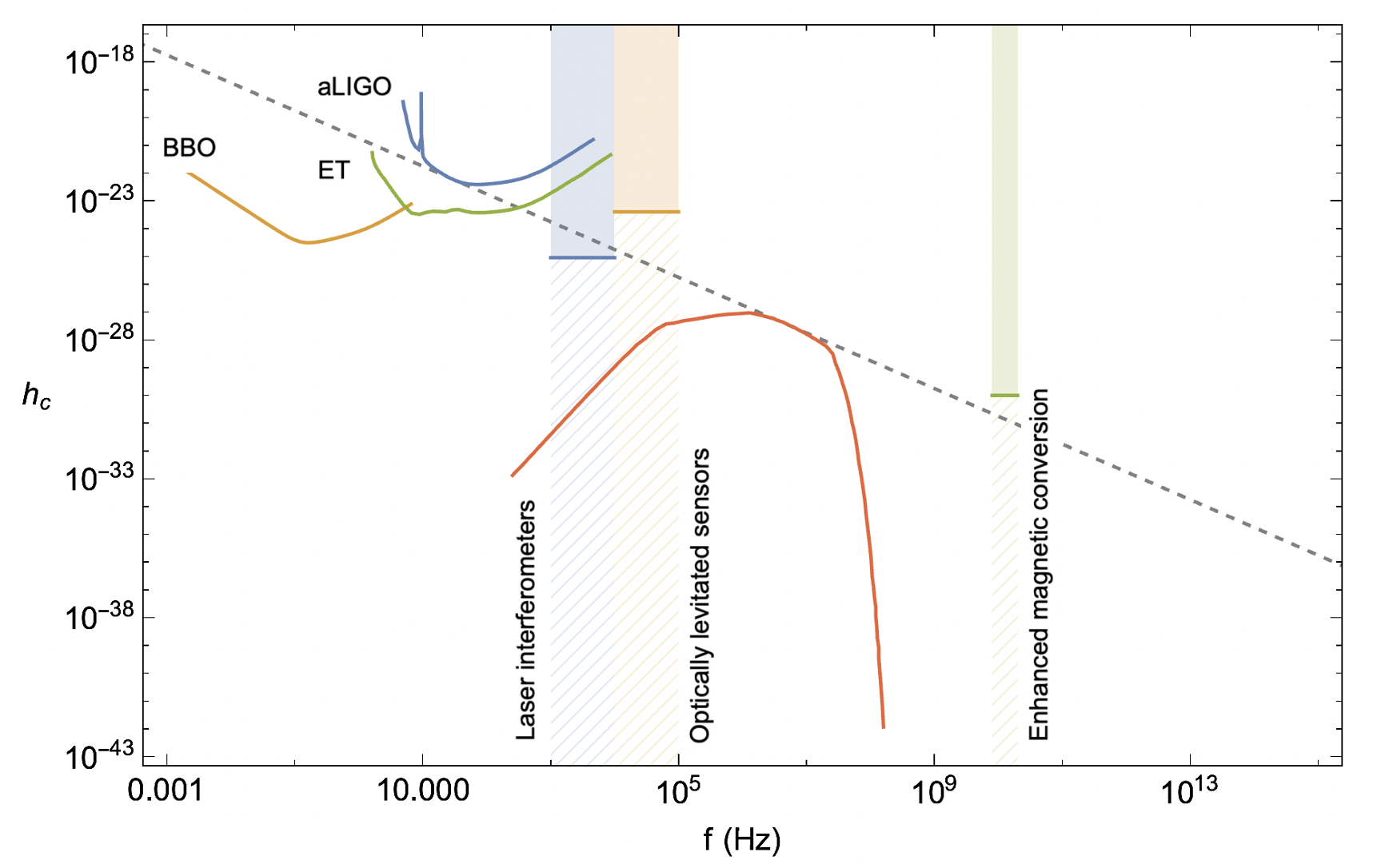}
\caption{\noindent Gravitational wave prediction (in terms of the dimensionless characteristic strain $h_c$) for an ``optimal'' scenario for experimental detection --- i.e. the lowest frequency signal with maximal amplitude consistent with $\Delta N_{\rm eff}$ constraints (grey dashed line). The corresponding parameter values are given in Eq.~(\ref{eq:sampleset}). Superposed are several current and proposed gravitational wave detectors and their projected sensitivities to a stochastic gravitational wave background. These include the ground-based observatories Advanced LIGO (aLIGO) and Einstein Telescope (ET) and the space-based Big Bang Observer (BBO) (sensitivities taken from \cite{Moore:2014}), as well as several prospective high-frequency gravitational wave detection technologies: laser interferometers, optically levitated sensors, and enhanced magnetic conversion (sensitivities taken from \cite{Aggarwal:2020}).}
\label{fig:experiments}
\end{figure}

From these trends, we are able to identify the ``best-case'' scenario illustrated in Fig.~\ref{fig:experiments}. The parameter values which yield the lowest peak frequency at the maximal amplitude still consistent with the $\Delta N_{\rm eff}$ bound are
\begin{equation}\label{eq:sampleset}
    n=2,\,\beta = 1,\, T_{\rm re}=16.5 \,\text{GeV},\, t_i = 10^{-30} \,\text{s},\,M_*=10^3\,\unit{TeV},\, M=10M_* \,.
\end{equation}
Note that the large degeneracy in parameter space means this is only one of many possible sets which could yield this result. 
While the tail of the signal extends down to hundreds of Hz, well within the range where Advanced LIGO (aLIGO) and Einstein Telescope (ET) are sensitive to a stochastic background \cite{Moore:2014}, the amplitude is almost 10 orders of magnitude too small for detection. We note, however, that the peak frequency plateaus in the sub-MHz range accessible to planned high-frequency gravitational wave detectors, making this scenario a target for detection once their sensitivity exceeds $\Delta N_{\rm eff}$ bounds.

\section{Conclusions}
\label{sec:conclusions}

Relic gravitational waves from the Hawking evaporation of tiny primordial black holes are a generic prediction of many early universe scenarios beyond the standard cosmological paradigm. For light primordial black holes evaporating before BBN, however, the resultant gravitational wave spectra generically peak at ultra-high frequencies $\sim M_{\rm Pl}^2/M$. As discussed in \cite{Ireland:2023}, these signals remain out of reach of even the most optimistic proposed high frequency detectors, such that the only way to probe these scenarios is through integral bounds on the energy density contained in gravitational waves. This is sufficient for putting broad constraints on such scenarios, but does not provide any detailed information.

This motivated the consideration of primordial black hole evaporation in the context of theories with large extra dimensions, in which the reduced value of the true bulk Planck scale $M_* \ll M_{\rm Pl}$ allows for lower peak frequencies $\sim \left( M_*/M \right)^{1/(n+1)} M_*$. 
Our ultimate goal was to make predictions for gravitational wave spectra from the Hawking evaporation of ultra-light black holes in such LED theories. To faithfully model black hole evaporation and evolution, we computed in detail the greybody factors for all particle species localized on the 4-dimensional ``brane'' as well as those propagating in the bulk. The resultant absorption coefficients are summarized in Table \ref{absorbcoefftable}. We were then able to make predictions for the gravitational wave signal. Figs.~\ref{fig:changingN} and \ref{fig:changingparameters} show the effect of changing the number of extra dimensions $n$, the bulk Planck scale $M_*$, the black hole mass $M$, the reheating temperature $T_{\rm re}$, and the formation time $t_i$. Figs.~\ref{fig:frequencycontours} and \ref{fig:amplitudecontours} explore this parameter space in more detail and present contour plots for the peak frequency and maximal amplitude, respectively. The trends established in these plots allowed us to identify an ``optimal'' scenario for experimental detection, featured in Fig.~\ref{fig:experiments}. 

As expected, we find that the lowest possible peak frequencies are obtained for models with very low bulk Planck scale $M_*$ and very light black holes with mass only slightly above $M_*$. The dependence on the number of extra dimensions is weaker, but increasing $n$ weakly pushes $f_{\rm peak}$ to slightly higher frequencies. Finally, the peak frequency is relatively insensitive to formation time and reheating temperature, though these affect the amplitude of the gravitational wave signal. We find that for the sample ``optimal'' parameter set of Eq.~(\ref{eq:sampleset}), the peak frequency plateaus in the sub-MHz range. This should be accessible to planned high-frequency gravitational wave detectors, once their sensitivities are improved to exceed the $\Delta N_{\rm eff}$ bounds. The tail of this spectrum extends into the $\sim 100$ Hz range, however in order for this tail to have an amplitude within sensitivity, the peak would violate the integral bound on the effective number of relativistic degrees of freedom $\Delta N_{\rm eff}$. Should high-frequency ($f\gtrsim$ kHz) observatories dip below the $\Delta N_{\rm eff}$ constraints, observation of evaporating extra dimensional black holes could be possible.

\section*{Acknowledgments}

S.P. and J.S. are supported in part by the U.S. Department of Energy grant number de-sc0010107.

\bibliographystyle{JHEP}
\bibliography{biblio}

\end{document}